\documentclass[]{article}
\makeatletter\if@twocolumn\PassOptionsToPackage{switch}{lineno}\else\fi\makeatother
\usepackage[utf8]{inputenc}
\usepackage{setspace}
\usepackage{amsfonts,amssymb,amsbsy,latexsym,amsmath,tabulary,graphicx,times,caption,fancyhdr,amsthm}
\usepackage{url,multirow,morefloats,floatflt,cancel,tfrupee}

\makeatletter

\AtBeginDocument{\@ifpackageloaded{textcomp}{}{\usepackage{textcomp}}}
\makeatother
\usepackage{colortbl}
\usepackage{xcolor}
\usepackage{pifont}
\usepackage[nointegrals]{wasysym}
\makeatletter

\def\mcWidth#1{\csname TY@F#1\endcsname+\tabcolsep}

\def\cAlignHack{\rightskip\@flushglue\leftskip\@flushglue\parindent\z@\parfillskip\z@skip}
\def\rAlignHack{\rightskip\z@skip\leftskip\@flushglue \parindent\z@\parfillskip\z@skip}

\@ifundefined{etal}{}{}

\usepackage{ifxetex}
\ifxetex\else\if@twocolumn\@ifpackageloaded{stfloats}{}{\usepackage{dblfloatfix}}\fi\fi

\AtBeginDocument{
\expandafter\ifx\csname eqalign\endcsname\relax
\def\eqalign#1{\null\vcenter{\def\\{\cr}\openup\jot\m@th
  \ialign{\strut$\displaystyle{##}$\hfil&$\displaystyle{{}##}$\hfil
      \crcr#1\crcr}}\,}
\fi
}

\AtBeginDocument{%
  \@ifpackageloaded{endfloat}%
   {\renewcommand\efloat@iwrite[1]{\immediate\expandafter\protected@write\csname efloat@post#1\endcsname{}}}{\newif\ifefloat@tables}%
}%

\def\BreakURLText#1{\@tfor\brk@tempa:=#1\do{\brk@tempa\hskip0pt}}
\let\lt=<
\let\gt=>
\def\processVert{\ifmmode|\else\textbar\fi}

\@ifundefined{subparagraph}{
\def\subparagraph{\@startsection{paragraph}{5}{2\parindent}{0ex plus 0.1ex minus 0.1ex}%
{0ex}{\normalfont\small\itshape}}%
}{}

\newcommand\role[1]{\unskip}
\newcommand\aucollab[1]{\unskip}
  
\@ifundefined{tsGraphicsScaleX}{\gdef\tsGraphicsScaleX{1}}{}
\@ifundefined{tsGraphicsScaleY}{\gdef\tsGraphicsScaleY{.9}}{}
\def\checkGraphicsWidth{\ifdim\Gin@nat@width>\linewidth
	\tsGraphicsScaleX\linewidth\else\Gin@nat@width\fi}

\def\checkGraphicsHeight{\ifdim\Gin@nat@height>.9\textheight
	\tsGraphicsScaleY\textheight\else\Gin@nat@height\fi}

\def\fixFloatSize#1{}
\let\ts@includegraphics\includegraphics

\def\inlinegraphic[#1]#2{{\edef\@tempa{#1}\edef\baseline@shift{\ifx\@tempa\@empty0\else#1\fi}\edef\tempZ{\the\numexpr(\numexpr(\baseline@shift*\f@size/100))}\protect\raisebox{\tempZ pt}{\ts@includegraphics{#2}}}}

\AtBeginDocument{\def\includegraphics{\@ifnextchar[{\ts@includegraphics}{\ts@includegraphics[width=\checkGraphicsWidth,height=\checkGraphicsHeight,keepaspectratio]}}}

\DeclareMathAlphabet{\mathpzc}{OT1}{pzc}{m}{it}

\def\URL#1#2{\@ifundefined{href}{#2}{\href{#1}{#2}}}

\def\UrlOrds{\do\*\do\-\do\~\do\'\do\"\do\-}%
\g@addto@macro{\UrlBreaks}{\UrlOrds}

\edef\fntEncoding{\f@encoding}

\makeatother

\newif\ifmultipleabstract\multipleabstractfalse%
\makeatletter

\def\wileyIndent{1pt}
\usepackage[paperheight=10in,paperwidth=6.5in,margin=2cm,headsep=.5cm,top=2.5cm,headheight=1cm]{geometry}

\renewenvironment{abstract}
{\vspace*{-1pc}\trivlist\item[]\leftskip\wileyIndent\hrulefill\par\vskip4pt\noindent\textbf{Summary}\mbox{\null}\\}{\par\noindent\hrulefill\endtrivlist}

\usepackage[]{footmisc}

\def\author#1{\gdef\@author{\hskip-\dimexpr(\tabcolsep)\hskip\wileyIndent\parbox{\dimexpr\textwidth-\wileyIndent}{\centering\bfseries#1}}}

\def\title#1{\linespread{1}\gdef\@title{\centering\bfseries\ifx\@articleType\@empty\else\@articleType\\\fi#1}}

\let\@articleType\@empty \def\articletype#1{\gdef\@articleType{{\normalfont\itshape#1}}}

 \def\audegree#1{}

\date{}

\makeatother

\usepackage[T1]{fontenc}
\makeatother
\usepackage[authoryear,round]{natbib}

\def\thanksspace{{\phantom{\textsuperscript{\thefootnote}}}}

\newcommand{\R}{\mathbb{R}}
\newcommand{\M}{\mathbb{M}}

\newtheorem{theorem}{Theorem}

\usepackage{xcolor}

\begin{document}
\title{Nonparametric testing of the dependence structure among points-marks-covariates in spatial point patterns}
\author{Ji\v{r}\'{\i}~Dvo\v{r}\'{a}k\textsuperscript{1}\thanks{Corresponding author. E-mail: dvorak@karlin.mff.cuni.cz.}{\thanksspace}, 
Tom\'{a}\v{s}~Mrkvi\v{c}ka\textsuperscript{2}, 
Jorge~Mateu\textsuperscript{3}\space 
and Jonatan~A.~Gonz\'{a}lez\textsuperscript{3}~\\[-3pt]\normalsize\normalfont  \itshape ~\\
\textsuperscript{1}{Department of Probability and Mathematical Statistics, Faculty of Mathematics and Physics, Charles University, Sokolovsk\' a 83, 186 75 Prague, Czech Republic}~\\
\textsuperscript{2}{Department of Applied Mathematics and Informatics, Faculty of Economics, University of South Bohemia, Studentsk{\'a} 13, 370 05 \v{C}esk\'e Bud\v{e}jovice, Czech Republic}~\\
\textsuperscript{3}{Department of Mathematics, University Jaume I, E-12071 Castell\'on, Spain}}

\def\RunningHead{}\def\RunningAuthor{Dvo\v{r}\'{a}k \MakeLowercase{\textit{et al.}} }

\maketitle 

\begin{abstract}
We investigate testing of the hypothesis of independence between a covariate and the marks in a marked point process. It would be rather straightforward if the (unmarked) point process were independent of the covariate and the marks. In practice, however, such an assumption is questionable and possible dependence between the point process and the covariate or the marks may lead to incorrect conclusions. Therefore, we propose to investigate the complete dependence structure in the triangle points--marks--covariates together. We take advantage of the recent development of the nonparametric random shift methods, namely the new variance correction approach, and propose tests of the null hypothesis of independence between the marks and the covariate and between the points and the covariate. We present a detailed simulation study showing the performance of the methods and provide two theorems establishing the appropriate form of the correction factors for the variance correction. Finally, we illustrate the use of the proposed methods in two real applications.

\def\keywordstitle{Keywords}

\smallskip\noindent\textbf{Keywords: }{Covariate, Hypothesis testing, Independence, Marked point process, Nonparametric inference}
\end{abstract}

\section{Introduction}

\begin{figure}[tb]
    \centering
    \includegraphics[width=\textwidth]{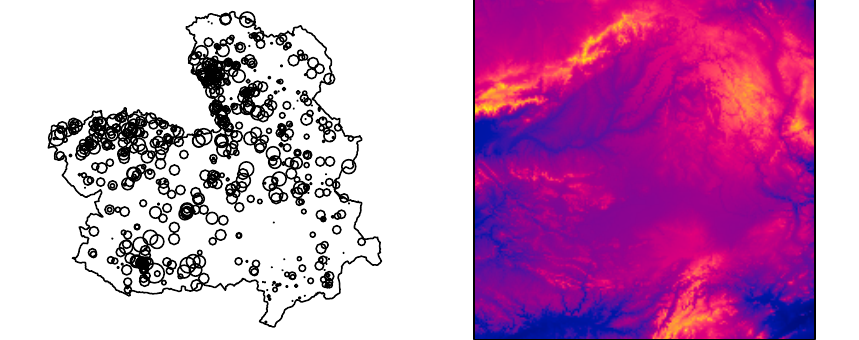}
    \caption{Castilla-La Mancha forest fires. Left: centroids of the individual forest fires (given by the centers of the discs) and their respective total area burned (logarithm of which is proportional to the discs' radii). Right: the terrain elevation, treated as the covariate in this example. For more details see Section~\ref{sec:clm}.}
    \label{fig:clm_intro}
\end{figure}

A marked point process is a model for a random collection of points in space to which a random mark is attached. For example, the left part of Figure~\ref{fig:clm_intro} shows the locations of a set of forest fires together with their total area burned, i.e. the marks. A natural question is whether the marks correlate with a given spatial covariate. The right panel of Figure~\ref{fig:clm_intro} shows such a covariate -- the terrain elevation in the given area. Detailed inspection (described in Section~\ref{sec:clm}) suggests that in places with high values of the covariate, smaller values of the marks are likely to occur,
and vice versa. Hence, one might be interested in formulating the null hypothesis of independence between the marks and the covariate and look for a formal statistical test. 

This question seems to be an objective of the geostatistical field, but this is true only when the marks are independent of the point locations. This marking model is then usually called the geostatistical marking model \citep{IllianEtal2008}. Similarly, the point locations may be influenced by the covariate, introducing the preferential sampling issues \citep{Diggle2010a}, meaning that the process of sampling the locations is not (stochastically) independent of the sampled phenomenon. This may lead to incorrect conclusions about the possible dependence between the marks and the covariate. The absence of dependence between the sampling locations and the covariate is then referred to as non-preferential sampling.

The problem of testing independence between marks and a covariate can be treated as testing independence between a pair of random fields, using e.g. Moran eigenvalue regression \citep{DrayEtal2006} or spatial cross-correlation \citep{Chen2015} only if the points are independent of the covariate, and under the geostatistical marking model. These strict requirements make it desirable to investigate the complete structure of dependence between points (P), marks (M), and a covariate (C) to make sure the inference about the possible dependence between marks and a covariate is valid.

In the framework of point processes, the question of dependence between marks and covariates has been studied so far  in a parametric setting where the whole model must be specified. \citet{IllianEtal2013} used the INLA approach to explore the dependence between a covariate and the whole point process. \citet{BernardoEtal1998} used a Bayesian hierarchical spatial regression model for the same purpose. \citet{MyllymakiPenttinen2009} studied the intensity-dependent marking models and used a Bayesian approach to test the independence of the mark values on the intensity of points. On the other hand, this paper aims to study the dependence between marks and a covariate in a nonparametric fashion, without imposing any model assumptions.

The problem described above considers both a continuous mark distribution and a continuous covariate. However, the approach proposed in this paper is also applicable for categorical (discrete) mark distributions, categorical (discrete) covariates, or even multiple covariates. In this paper, we assume, for ease of presentation, that we have only one covariate. We assume that its values are available at any point of the observation window, either observed, at least on a fine pixel grid (e.g. altitude, distance from a geological fault) or extrapolated from given measurements through kriging, conditional simulation, or other procedures (e.g. mineral content in the soil, level of air pollution). The case of multiple covariates is briefly discussed in Section~\ref{sec:multiple}.

Our proposed nonparametric tests of independence between marks and a covariate (and points and a covariate) are based on random shifts. The random shift approach takes advantage of Monte Carlo testing and was recommended for testing the independence hypothesis in the geostatistical context by \citet{UptonFingleton1985} and popularized by \citet{DaleFortin2002}. In the point process literature, the random shift approach is used for testing independence between a pair of point processes \citep{Lotwick1982,GrabarnikEtal2011}. 

The random shift tests are based on a specific strategy for producing Monte Carlo replications \citep{Lotwick1982}. To break any possible dependence structure between a pair of spatial processes (such as a marked point process and a covariate), one of the processes is kept fixed while the other one is shifted by a random vector. Different versions of the random shift test are available, using different ways to deal with the part of the data that is shifted outside the observation window $W$. In this paper, we focus on two of them, the well-established torus correction and the variance correction introduced in \citet{MrkvickaEtAl2020} for testing both the independence of two random fields and the independence of two point processes observed in the same spatial domain.

The torus correction approach \citep{Lotwick1982} makes the shifts respecting the toroidal geometry induced by identifying the opposite edges of the rectangular observation window.
After a shift with torus correction, abrupt changes or artificial modifications in the correlation structure of the data appear at places where originally distant parts of the data are glued together. This, in turn, introduces the liberality of the test. To compensate for the liberality, \cite{MrkvickaEtAl2020} proposed a variance correction strategy based on dropping out the part of the data shifted outside $W$. In this way, no unnecessary changes are created in the correlation structure. On the other hand, the amount of data used for computing the test statistic values in the Monte Carlo test changes and their means and variances are possibly different, requiring standardization to get closer to exchangeability. Details are given in \citet[Sec. 2.1.3]{MrkvickaEtAl2020}. The variance correction version of the random shift test can also be used for irregular windows, making its practical applicability higher compared to the torus correction.

As mentioned above, to investigate the relationship between marks and a covariate, their relationships with the point locations need to be considered.
The hypothesis of independence between marks and point locations can be  tested by the methods introduced in \citet{SchlatherEtal2004} or \citet{Guan2006}. A test of this hypothesis will be denoted a P-M test in the following. Similarly, the hypothesis of independence between point locations and the covariate can be tested using the same tests, along with other possibilities such as \citeauthor{Berman1986}'s tests using the Poisson assumption \citep{Berman1986} or the parametric approach of \cite{WaagepetersenGuan2009}. Our paper proposes a new, fully nonparametric test of independence between a point process and a covariate, based on the random shift approach. A test of this hypothesis will be denoted a P-C test in the following.

The core of the paper lies in nonparametric testing of independence between marks and a covariate. Such a test will be denoted PM-C test in the following, to stress the fact that the points and marks are inseparable, and, when the points are shifted in the random shift procedures, they are always shifted together with their respective marks. This property implies that rejection in the PM-C test may be caused by dependence between marks and the covariate, but also by the possible dependence between marks and points and points and the covariate, respectively. For this reason, we believe that a PM-C test should always be accompanied by a P-C and a P-M test that help interpret the outcome of the PM-C test.

We propose several versions of the PM-C test based on the random shift approach, using different test statistics and either the torus correction or the variance correction. We investigate the performance of the tests in an extensive simulation study and illustrate the practical use of the methods in two real data examples. The potential benefit of the proposed tests lies in suggesting which covariates are relevant in the given application and should be accounted for in the following steps of the statistical analysis, without specifying a parametric model at this stage.

The paper is organized as follows. Section~\ref{sec:background} contains the necessary background material for the random shift tests. The proposed methods are described in Section~\ref{sec:methods}. The performance of the methods is assessed in Section~\ref{sec:simulations} by simulations, and their practical use is illustrated in Section~\ref{sec:applications} for two real datasets. Some concluding remarks are given in Section~\ref{sec:conclusions}. The Appendix contains the proofs of two theorems from Section~\ref{sec:methods} and a~numerical illustration of the order of variance discussed in the theorems.

\section{Background and notation}\label{sec:background}

Let $\Psi$ be a stationary marked point process on $\R^2$ with mark space $\M$, and let $\Phi$ be the corresponding unmarked point process. We assume throughout the paper that $\M=\R$ unless explicitly stated otherwise.
Let $Z(u), u \in \R^2,$ denote a stationary random field representing the covariate. Assume that both $\Psi$ and  $Z$ are observed in a compact observation window $W \subset \R^2$ and that the values $Z(u)$ are available for all $u \in W$.

The tests proposed in Section~\ref{sec:methods} are Monte Carlo tests, comparing the value of the test statistic $T_0$ computed from the observed data with a set of values $T_1, \ldots, T_N$ obtained from a set of replications. These are produced so that they follow the null hypothesis $H_0$, and hence the values $T_1, \ldots, T_N$ approximate the distribution of $T_0$ under $H_0$. The test outcome is then determined based on how typical or extreme the value of $T_0$ is among $T_1, \ldots, T_N$. The test statistic can be univariate, multivariate, or functional.

Monte Carlo tests rely on the exchangeability of $(T_0, T_1, \ldots, T_N)$ in order to be exact. This is easily fulfilled if the replications are independent of the observed data and are independent, identically distributed under $H_0$. Whenever independent replications are not available, as is often the case when nonparametric methods are employed, a replication strategy must be chosen to achieve exchangeability. If this is not possible, the aim is to make the deviation from exchangeability as small as possible. For example, the often used Freedmann and Lane permutation scheme \citep{FreedmanLane1983} in general linear models also does not achieve exchangeability, but it performs better than alternative approaches, and therefore it is usually recommended.

Since the exchangeability of $(T_0, T_1, \ldots, T_N)$ implies a uniform distribution of the p-value, potential deviations from the exchangeability can be detected in carefully designed si\-mu\-la\-tion experiments studying the size of the test. Different tests of the same hypothesis can be compared in this way, with more severe deviations from the uniform distribution of p-values implying more severe deviations from exchangeability and hence inferiority of the test.

The Monte Carlo tests considered in this paper differ in many aspects, but they all fit into one of the following classes of tests, depending on the replication strategy.

\subsection{Simulation-based tests}

The most straightforward approach is to fit a parametric model to the observed data and obtain the required replications as independent simulations from the fitted model. This approach has the drawbacks common to all parametric procedures: possible issues with model misspecification or possibly poor parameter estimation. Also, the tests often use a composite null hypothesis, which introduces conservativeness. Finally, the models in stochastic geometry are often complicated, and it can be very time-consuming to simulate the required number of realizations. The test proposed in \citet{SchlatherEtal2004} for testing the geostatistical marking model fits into this class of tests.

\subsection{Random shift approach}

One of the popular nonparametric strategies for testing in spatial statistics is the random shift approach. It is useful when two random objects $\Psi_1, \Psi_2$ (random fields, point processes, random sets, etc.), possibly influencing each other, are observed. The null hypothesis of their independence can be tested in a Monte Carlo fashion by producing the replications by randomly shifting one of the objects while keeping the other fixed. The shifts are intended to break any possible dependence between $\Psi_1$ and $\Psi_2$ and produce replications where $\Psi_1,\Psi_2$ are (nearly) independent. Approximately exchangeable values of the test statistic can be obtained by using many (randomly chosen) shifts. This enables the application of the Monte Carlo test in an approximate manner. The appropriateness of this approach should be justified for a given problem by a simulation study.

Several variants of the random shift approach are available, which differ in how they treat the part of data shifted outside the observation window. Below, we discuss the classical, well-established torus correction approach and the variance correction approach that turned out to be very efficient in a recent study in \citet{MrkvickaEtAl2020}. In the current paper, the two random objects to be shifted against each other are the marked point process $\Psi=\Psi_1$ and the random field $Z=\Psi_2$.

\subsubsection{Torus correction}\label{sec:torus_correction}

Assuming the observation window $W$ is rectangular, its opposite sides can be identified, inducing a toroidal geometry on $W$. In our setting, the marked point process $\Psi$ is shifted by a random vector $v$, respecting the toroidal geometry. Hence, no points leave $W$, keeping the number of observed points the same for the original data and all the replications. The torus correction introduces a crack in the correlation structure of the marked point process, causing the liberality reported by \citet{FortinPayette2002} and \citet{MrkvickaEtAl2020}.

\subsubsection{Variance correction}\label{sec:variance_correction}

Both the requirement of rectangular observation window and the liberality of torus correction are serious drawbacks, motivating the introduction of the variance correction approach in \citet{MrkvickaEtAl2020}. In this strategy, the points that leave $W$ after the shift are discarded. In this way, no unwanted artifacts are introduced in the correlation structure of the data, but the number of points in each replication differs. This, in turn, influences the distribution of the test statistic, most notably the variance. Hence, the first two moments of the test statistic need to be standardized, bringing the vector $(T_0, T_1, \ldots, T_N)$ closer to exchangeability.

Formally, the values of the test statistic $T_0, T_1, \ldots, T_N$ are computed from the original and shifted data. The overall mean $\overline{T} = \frac{1}{N+1} \sum_{i=0}^N T_i$ is subtracted (assuming that the expectation of the test statistic is not affected by the random shifts) and the variance is standardized by dividing by $\sqrt{\mathrm{var} (T_i)}$:
\begin{align*}
    S_i = \left( T_i - \overline{T} \right) / \sqrt{\mathrm{var} (T_i)}, \quad i = 0, \ldots, N.
\end{align*}
The transformed values $S_0, S_1, \ldots, S_N$ are then used to determine the p-value of the test in the classic Monte Carlo fashion.

Several ways to determine $\mathrm{var} (T_i)$ are available, depending on the particular choice of the test statistic \citep{MrkvickaEtAl2020}. Here, we consider the approach based on the asymptotic order of $\mathrm{var} (T)$ as a function of the number of observed points; some details are given below.

\section{Methods}\label{sec:methods}
In this section, we describe the tests that can be used to assess the dependence structure in the triangle points--marks--covariates.

\subsection{P-M test}\label{subsec:PM}
The null hypothesis of independence between marks and points in a marked point process needs to be appropriately specified, since the mark values are only defined at the points of the pattern. There are two natural ways to specify the null hypothesis \citep[Sec. 7.5]{IllianEtal2008}, the independent marking model and the geostatistical marking model. In this paper, we focus on the latter, where a random field $Z_M$ gives the mark values sampled at the points of a point process $\Phi$, independent of $Z_M$. This model covers the more general case of correlated marks.

\citet{SchlatherEtal2004} introduce characteristics $E(t)$ and $V(t)$ representing the conditional expectation and conditional variance of a mark, respectively, given that another point of the process is at distance $t$. Under the null hypothesis of the geostatistical marking model, these characteristics are constant. The paper suggests a Monte Carlo test of the null hypothesis based on simulations of Gaussian random fields with the correlation structure estimated from the data. This assumption of a normal distribution is important (but questionable for many real datasets), and hence the observed data must be first transformed marginally to normal variables, using the empirical distribution function.

Based on the simulation experiments reported in \citet{SchlatherEtal2004}, we choose to work only with the function $E(t)$ and construct the test statistic using the $l_2$-norm (with constant weights) of the difference between the estimated function $\widehat{E}(t)$ and the constant function $\widehat{E}(0)$, the mean observed mark value.

To avoid the normality assumption and the need to simulate from a fitted model, \citet{GuanAfshartous2007} propose a method based on subsampling to estimate a covariance matrix needed for computing their test statistic. In practice, this requires a large dataset, and we do not consider this method here. \citet[Sec. 7.5.3]{IllianEtal2008} consider a test of the geostatistical marking model using the mark-weighted $L$-function and random permutations of the marks among points. However, this approach breaks the mark correlation structure, causing extreme liberality of the test, and we do not consider it here.

\citet{Guan2006} studies in detail the independence between points and marks and provides both graphical tools and formal tests for assessing the possible dependence. One of the tests is based on comparing the empirical mark-weighted $K$-function to the empirical $K$-function of the unmarked point process and relies on the test statistic being asymptotically $\chi_1^2$ distributed.
This test performed well in simulations carried out in \cite{Guan2006} and we compare it here with the \cite{SchlatherEtal2004} test.

\subsection{P-C test}\label{subsec:PC}

When the covariate values are observed only at the points of the process, the independence between the points and the covariate can also be tested by the methods proposed for testing the geostatistical marking model. As an example, we consider here again the test by \citeauthor{SchlatherEtal2004}, as also suggested in \cite{Diggle2010}, and \cite{Guan2006} test. However, our main aim is to investigate the situation where the covariate values are available everywhere in the observation window and to propose nonparametric tests that fully exploit the available covariate information.

Our tests are based on the random shift approach either with the torus correction or the variance correction described in Sections~\ref{sec:torus_correction} and \ref{sec:variance_correction}. Assuming that the covariate has numeric values, the test statistic is the sample mean of the covariate values observed at the point pattern locations:
\begin{align}\label{eq:PCstat}
    T = \frac{1}{\Phi (W)} \sum_{x \in \Phi \cap W} Z(x).
\end{align}
The choice of the test statistic is motivated by the most natural scenario, where the points of the process are more likely to appear in locations with high (or low) covariate values. The test statistic $T$ is able to capture this type of dependence, attaining a higher (or lower) value for the observed data than for the shifted data, resulting in a test with high power in this setting. If one has a different alternative hypothesis in mind (more complicated type of dependence between the points and the covariate), another test statistic should be used.

We remark here that the simple point process $\Phi$ can be equivalently seen as a random locally finite set (justifying the set notation $\Phi \cap W$ in the formula above, denoting the set of points of the process $\Phi$ in the set $W$) or a random locally finite counting measure (justifying the measure notation $\Phi(W)$ in the formula above, denoting the number of points of $\Phi$ in $W$). Using both types of notation is quite common in the point process literature, leading to more comprehensible arguments and formulas.

The asymptotic order of the variance of the sample mean is $1/n$ when computed from a sample of $n$ i.i.d. observations. In our case, the following theorem gives, for a stationary point process $\Phi$ with intensity $\lambda$, that ${\mathrm{var}}(T) \approx 1/ ( \lambda |W| )$. Since the true intensity $\lambda$ is unknown, in practice we plug-in its estimator $\widehat{\lambda} = \Phi(W) / |W|$ and for variance correction we use the (estimated) correction term ${\mathrm{var}}(T) \approx 1/\Phi(W)$.

\begin{theorem}\label{T1}
Let $\Phi$ be a stationary point process in $\mathbb{R}^2$ with intensity $\lambda$ and pair-correlation function $g$, observed in the observation window $W$. Let $Z(u), u \in W$, be a centered stationary random field with finite second moments, independent of $\Phi$, having a non-negative covariance function $C$. Assume that there is a constant $R>0$ such that $C(u-v) = 0$ for $\| u - v \| > R$. Define the random variables $S,U$ in the following way:
\begin{align*}
    S = \sum_{x \in \Phi \cap W} Z(x), \qquad U = \Phi(W).
\end{align*}
Then there exist constants $0<C_1 \leq C_2 < \infty$, depending on the properties of $\Phi$ and $Z$ but not on $|W|$, such that
\begin{align*}
    C_1 \leq \frac{\mathrm{var}(S)}{\lambda |W|} \leq C_2.
\end{align*}
Moreover, the variance of $S/U$ can be approximated by
\begin{align*}
    \mathrm{var}\left( \frac{S}{U} \right) \approx \frac{\mathrm{var}(S)}{\lambda^2 |W|^2}
\end{align*}
and hence $\mathrm{var}(S/U)$ is of order $1/(\lambda|W|)$.
\end{theorem}

The proof is given in Appendix~\ref{app:proof1}. Note that the fraction $S/U = T$ gives the test statistic from \eqref{eq:PCstat}. Appropriate mixing conditions can replace the assumption of bounded support of the covariance function $C$.
We remark that the approximation above, based on the first-order Taylor expansion of the function $f(S,U) = S / U$, requires that $\mathbb{P}(U=0)=\mathbb{P}(\Phi(W)=0)=0$ which is, strictly speaking, not fulfilled for the usual point process models. Hence, one may consider $\tilde{U}=\Phi(W) + \epsilon$ for a small positive constant~$\epsilon$ instead, without changing the variance order. In Appendix~\ref{app:simulations_PC}, it is illustrated by simulation that the order of variance is indeed $1/{(\lambda|W|)}$.

Other test statistics can, of course, be used if there is a particular indication that other properties of the covariate $Z$ might influence the occurrence of points in $\Phi$. For example, the histogram (vector of counts of observations with values in disjoint intervals) can be used so that the whole distribution of the covariate values at points of the process is captured. In this case, the global envelope tests of \citet{MyllymakiEtal2017} can be used to perform the Monte Carlo test with a multivariate test statistic.

Naturally, if several covariates $Z_1, \ldots, Z_K$ are available, one can test them individually and bind the tests together by, say, Bonferroni correction. Alternatively, one can consider, as a test statistic, the vector of means $T=(T^1, \ldots, T^K)$, $T^i = \frac{1}{\Phi (W)} \sum_{x \in \Phi \cap W} Z_i(x), i = 1, \ldots, K$, and use, for example, the global envelope procedure \citep{MyllymakiEtal2017} to perform the test. This will be exemplified in Section~\ref{sec:bogota} when analyzing the Kennedy-Bogota crime dataset.

\subsection{PM-C test}\label{subsec:PMC}

When testing the independence between marks and a covariate, our starting point is choosing the sample covariance as the test statistic. This is motivated by the use of sample covariance for testing the independence between two random fields in \citet{MrkvickaEtAl2020}. Let $\{ (x_1,m_1), \ldots, (x_n,m_n) \}$ be the observed realization of the marked point process $\Psi$ and let $z(u), u \in W,$ denote the observed rea\-li\-za\-tion of the covariate $Z$. We define
\begin{align*}
    T_C = \frac{1}{n-1} \sum_{i=1}^n (m_i - \overline{m})(z(x_i) - \overline{z}),
\end{align*}
where 
\begin{align*}
    \overline{m} = \frac{1}{n} m_i, \qquad \overline{z} = \frac{1}{n} z(x_i)
\end{align*}
are the observed sample means. The asymptotic order of the variance of the sample covariance is $1/n$ when computed from a sample of $n$ i.i.d. observations. The same holds for the sample covariance of the values of two independent random fields observed at deterministic sampling locations \citep{MrkvickaEtAl2020}. Therefore, we use ${\mathrm{var}}(T_C) \approx 1/n$ for variance correction.

Although this choice of a test statistic is perfectly appropriate in situations where the unmarked point process of sampling locations is independent of the marks and the covariate, it may perform poorly in cases where there is a dependence between them. In the case of dependence between the point process and the covariate, the problem of preferential sampling occurs \citep{Diggle2010}. This term is used in the geostatistical literature to describe the general situation where the set of sampling points is not independent of the studied random field, e.g., when more samples are taken at the locations where high-grade ore is thought likely to be found. It has been reported that preferential sampling introduces bias into the estimation of the covariance structure of the random field \citep{Diggle2010}.

A similar issue also applies in our context: preferential sampling may introduce bias when estimating the covariance between marks and a covariate. Random shifts then violate preferential sampling, changing the distribution of the test statistic computed from the shifted distribution. This, in turn, damages exchangeability and the resulting test is far from exact. Whether the test would be conservative or liberal depends on the particular type of preferential sampling. The same issues can be caused by the dependence between the points and marks. However, we do not use the term ``preferential sampling'' in this case to avoid confusion.

This leads us to define different test statistics that will be less affected by the bias in the estimated covariance structure of the covariate and the marks, more specifically, less affected by the sample variance. We choose Pearson's correlation coefficient and, assuming no ties are present in the data, Kendall's rank correlation coefficient:
\begin{align*}
    T_P & = \frac{\sum_{i=1}^n (m_i - \overline{m})(z(x_i) - \overline{z})}{\sqrt{\sum_{i=1}^n (m_i - \overline{m})^2}\sqrt{\sum_{i=1}^n (z(x_i) - \overline{z})^2}}, \\
    T_K & = \frac{1}{n(n-1)} \sum_{i \neq j}\mathrm{sgn}(m_i - m_j) \mathrm{sgn}(z(x_i) - z(x_j)).
\end{align*}

The asymptotic order of the variance of $T_P$, when computed from a sample of $n$ i.i.d. observations, is $1/n$ \citep[p. 30]{van1998asymptotic}. The same holds for $T_K$ \citep[pp. 164--165]{van1998asymptotic}. Therefore, we use ${\mathrm{var}}(T_P), {\mathrm{var}}(T_K) \approx 1/n$ for variance correction. This is justified for $T_K$ by the following theorem which states that, for a stationary point process $\Phi$ with intensity $\lambda$, the variance of $T_K$ is of the order $1/ ( \lambda |W| )$. Since the true intensity $\lambda$ is unknown, in practice we plug-in its estimator $\widehat{\lambda} = \Phi(W) / |W|$ and for the variance correction the (estimated) correction term is ${\mathrm{var}}(T) \approx 1/\Phi(W)$.

\begin{theorem}\label{T2}
Let $\Psi$ be a stationary marked point process in $\mathbb{R}^2$, observed in the observation window $W$. Let $\Psi$ follow the geostatistical marking model, i.e. it is obtained by sampling the (random) mark field $Z_1$ at points of the unmarked point process $\Phi$. Assume that the product densities of $\Phi$ up to the fourth order exist and are bounded by finite positive constants both from above and from below. They will be denoted by $\lambda, \lambda_2, \lambda_3$ and $\lambda_4$ in the following.

Let the covariate be given by the random field $Z_2$ and let the random fields $Z_1, Z_2$ be independent, identically distributed, centered stationary Gaussian random fields with a non-negative covariance function $C$. Assume that there is a constant $R$ such that $C(u-v)=0$ for $\| u-v \| > R$. Furthermore, assume that there are constants $\delta > 0$ and $r > 0$ such that $C(u-v) \geq \delta$ for $\| u-v \| \leq r$. Define the random variables $S,U$ in the following way:
\begin{align*}
    S =  \sum_{x,y \in \Phi \cap W}^{\neq}\mathrm{sgn}(Z_1(x) - Z_1(y)) \, \mathrm{sgn}(Z_2(x) - Z_2(y)), \quad U = \Phi(W)(\Phi(W)-1).
\end{align*}
Then there exist constants $0<c_1 \leq c_2 < \infty$ and $0<d_1 \leq d_2 < \infty$, depending on the properties of $\Phi$ and $Z_1, Z_2$ but not on $|W|$, such that $\mathrm{var}(S) = A + B$ with
\begin{align*}
    c_1 \leq \frac{A}{\lambda^3 |W|^3} \leq c_2, \qquad d_1 \leq \frac{B}{\lambda^2 |W|^2} \leq d_2.
\end{align*}
Moreover, the variance of $S/U$ can be approximated by
\begin{align*}
    \mathrm{var}\left( \frac{S}{U} \right) \approx \frac{\mathrm{var}(S)}{(\mathbb{E}U)^2}
\end{align*}
with $u_1 \leq \mathbb{E}U / (\lambda |W|)^2 \leq u_2$ for some finite positive constants $u_1, u_2$. Therefore, $\mathrm{var}(S/U)$ is of order $1/(\lambda|W|)$.
\end{theorem}

The proof is given in Appendix~\ref{app:proof2}. Note that the fraction $S/U = T$ gives the test statistic $T_K$. The assumption of bounded support of the covariance function $C$ can be replaced by appropriate mixing conditions. In Theorem~\ref{T2} the assumption of $Z_1, Z_2$ being equally distributed may be weakened to the two random fields having different variances, i.e. the two covariance functions being proportional to each other, without any changes to the proof apart from introducing more complicated notation. In Appendix~\ref{app:simulations_PMC} it is illustrated by simulation that the order of variance is $1/{(\lambda|W|)}$ both in the case of equal covariance functions and in the case of unequal covariance functions, one of them even having unbounded support.

In case of analyzing categorical marks with mark space $\mathbb{M}=\{1,\ldots,M\}$, i.e. a~multi-type point process, one can consider as the test statistic the vector of differences between the means of the covariate values in the individual component processes. More formally, denoting $T^i = \frac{1}{\Phi_i (W)} \sum_{x \in \Phi_i \cap W} Z(x), i = 1, \ldots, M$, where $\Phi_i$ is the component process of points with marks ``i'', the test statistic would be the vector $T=(T^1-T^2, T^1-T^3, \ldots, T^{M-1}-T^M)$.  The global envelope procedure can then be used to perform the test. This will be exemplified in Section~\ref{sec:bogota} when analyzing the Kennedy-Bogota crime dataset, also in the case of testing multiple covariates at once.

In case of analyzing a categorical covariate and continuous marks, the categorical covariate divides the window into subwindows. The vector of differences of mean mark values in each pair of subwindows can be considered as a test statistic, similarly to the previous paragraph. Also, the case of categorical marks (multi-type point process) and a categorical covariate could be solved by a test statistic counting the number of points each type in every subwindow, in a contingency table manner. These cases deserve further investigation, which is outside the scope of this paper.

\section{Simulation experiments}\label{sec:simulations}

To assess the performance of the tests proposed in this paper, we conducted a simulation study with models covering all combinations of presence/absence of dependence between points and marks, points and covariate, marks and covariate. We present an overview study in Section~\ref{sec:overview}, followed by detailed studies focusing on the influence of preferential sampling and dependent marking in Sections~\ref{sec:preferentialSampling} and \ref{sec:dependentMarking}, respectively. We also include a small experiment with multiple covariates in Section~\ref{sec:multiple}.

In all simulation experiments described below, the following choices were made:
\begin{itemize}
    \item The random fields $Z_1, Z_2, \ldots$ are independent centered unit variance Gaussian random fields with the isotropic correlation function $c(r) = \exp\{-5r\}, r \geq 0$.
    \item To simplify the notation, we denote $a=1/\sqrt{2}$. Then we have, e.g., that $aZ_1 + aZ_2$ is a centered Gaussian random field with unit variance.
    \item The point process models are stationary log-Gaussian Cox processes with intensity $\lambda = \exp\{ 5 \} \doteq 148$, with the exception of models M10 and M12. However, these two models are constructed so that they are stationary with intensity $\exp\{5\}$, too.
    \item The marks are obtained by sampling a random field at the point process locations (i.e. the geostatistical marking model in cases where the point process and the mark field are independent).
    \item The mark fields and the covariates are centered unit variance Gaussian random fields, obtained by a linear combination of different $Z_i$s, again with the exception of models M10 and M12.
    \item The observation window is the unit square $[0,1]^2$.
    \item For each test based on random shifts, 999 independent random shifts are performed to obtain the Monte Carlo replications.
    \item The tests from \cite{SchlatherEtal2004} are based on 99 independent simulations from the fitted model. Note that in this study, the assumption of Gaussian distribution of the mark field or the covariate \citep{SchlatherEtal2004} is fulfilled, and the use of the tests is justified.
    \item \cite{Guan2006} test is based on the mark-weighted $K$-function with an upper bound $R=0.1$.
    \item  All tests are performed on the fixed 5\% significance level, and $5\,000$ independent realizations from each model are used. Fractions of rejection are reported in the tables of results below.
\end{itemize}

\subsection{Overview study}\label{sec:overview}

In this simulation experiment we consider eight models, denoted M1 to M8, covering all combinations of absence/presence of dependence between the three model components: points, marks, covariate. For easier interpretation of the results, the marginal distribution of the unmarked point process is the same for all eight models, and the same also holds for the marginal distribution of the mark field and the covariate. The detailed structure of the models is given in Table~\ref{tab:models_overall}, specifying the driving intensity function of the LGCPs and describing the mark fields and the covariates.

\begin{table}
\renewcommand{\arraystretch}{1.3}
\begin{tabular}{|l|l|c|c|c|}
 \hline
  & Structure & Points & Marks & Covariate \\ 
 \hline\hline
  M1 & P--M--C--P & $\exp\{4.5 + Z_1(u)\}$ & $Z_2(u)$ & $Z_3(u)$  \\ \hline
  M2 & P--M+C--P & $\exp\{4.5 + Z_1(u)\}$ & $a Z_2(u) + a Z_3(u)$ & $a Z_2(u) + a Z_4(u)$ \\ \hline
  M3 & P--M--C+P & $\exp\{4.5 + a Z_1(u) + a Z_3(u)\}$ & $Z_2(u)$ & $Z_3(u)$  \\ \hline
  M4 & P+M--C--P & $\exp\{4.5 + a Z_1(u) + a Z_2(u)\}$ & $Z_2(u)$ & $Z_3(u)$  \\ \hline
  M5 & P+M--C+P & $\exp\{4.5 + a Z_2(u) + a Z_3(u)\}$ & $Z_2(u)$ & $Z_3(u)$  \\ \hline
  M6 & P+M+C--P & $\exp\{4.5 + a Z_1(u) + a Z_2(u)\}$ & $a Z_2(u) + a Z_4(u)$ & $a Z_3(u) + a Z_4(u)$ \\ \hline
  M7 & P--M+C+P & $\exp\{4.5 + a Z_1(u) + a Z_3(u)\}$ & $a Z_2(u) + a Z_4(u)$ & $a Z_3(u) + a Z_4(u)$ \\ \hline
 M8 & P+M+C+P & $\exp\{4.5 + a Z_1(u) + a Z_2(u)\}$ & $a Z_1(u) + a Z_3(u)$ & $a Z_2(u) + a Z_3(u)$ \\ \hline 
\end{tabular}
\caption{Models for the overall study. For clarity, in the second column the true dependence structure of the point locations (P), marks (M) and the covariate (C) is given for each model. The ``+'' sign indicates dependence between the given pair of objects, the ``-'' sign indicates independence.}
\label{tab:models_overall}
\end{table}

\begin{table}
\small
\renewcommand{\arraystretch}{1.2}
\begin{tabular}{|l|c|c|c|c|c|c|c|c|}
 \hline
  & M1 & M2 & M3 & M4 & M5 & M6 & M7 & M8 \\ 
 \hline\hline
 P-M (Schlather)   & {\bf 0.060} & {\bf 0.064} & {\bf 0.064} & {\bf 0.507} & {\bf 0.493} & {\bf 0.266} & {\bf 0.059} & {\bf 0.249} \\ \hline
 P-M (Guan)        & 0.038 & {\bf 0.036} & 0.033 & 0.489 & 0.471 & 0.234 & 0.030 & 0.225 \\ \hline \hline
 
 P-C (torus)       & 0.068 & 0.067 & {\bf 0.982} & 0.066 & {\bf 0.982} & 0.073 & {\bf 0.804} & {\bf 0.791} \\ \hline
 P-C (variance)       & {\bf 0.041} & 0.036 & 0.917 & {\bf 0.043} & 0.919 & {\bf 0.042} & 0.617 & 0.613 \\ \hline
 P-C (Schlather)   & {\bf 0.059} & {\bf 0.060} & 0.500 & 0.072 & 0.516 & 0.059 & 0.264 & 0.259 \\ \hline
 P-C (Guan)        & 0.031 & 0.034 & 0.482 & 0.038 & 0.489 & 0.036 & 0.228 & 0.215 \\ \hline \hline
 PM-C (torus, cov) & 0.061 & 0.785 & 0.075 & 0.074 & 0.071 & 0.777 & 0.784 & 0.759 \\ \hline
 PM-C (variance, cov) & 0.059 & 0.799 & 0.065 & 0.065 & 0.069 & 0.801 & 0.795 & 0.771 \\ \hline
 PM-C (torus, Pea)    & 0.067 & 0.826 & 0.076 & 0.074 & 0.077 & 0.823 & 0.824 & 0.799 \\ \hline
 PM-C (variance, Pea) & 0.053 & {\bf 0.835} & 0.056 & 0.058 & 0.061 & {\bf 0.834} & {\bf 0.831} & {\bf 0.814} \\ \hline
 PM-C (torus, Ken)    & 0.062 & 0.794 & 0.072 & 0.068 & 0.070 & 0.793 & 0.788 & 0.771 \\ \hline
 PM-C (variance, Ken) & {\bf 0.050} & 0.804 & {\bf 0.055} & {\bf 0.055} & {\bf 0.057} & 0.799 & 0.790 & 0.772 \\ \hline
\end{tabular}
\caption{Results of the overall study: fraction of rejections of the corresponding null hypothesis. Results based on $5\,000$ independent replications. The confidence interval for the rejection rate based on the binomial distribution for the nominal significance level $0.05$ is [0.044,0.056].}
\label{tab:results_overall}
\end{table}

The results are given in Table~\ref{tab:results_overall}, showing the fraction of rejections of the individual tests for the given models. Values corresponding to the most successful test for each model are printed in bold, individually for the three types of tests (P-M, P-C or PM-C). This means that the test that matches the nominal significance level the best is highlighted where the null hypothesis holds, or the test with the highest power is highlighted where the null hypothesis does not hold.

\subsubsection*{Overview study: P-M tests}

We observe that the \cite{SchlatherEtal2004} test of independence between points and marks (P-M test) is slightly liberal under the null hypothesis (models M1, M2, M3, and M7). This is caused by the small bias in the estimation of the mark field parameters, which is caused by the fact that the mark field is sampled not in a uniformly chosen set of locations but in the points of a clustered point process. This is similar to the effect of the clustered sampling design reported by \cite{Diggle2010}.

The test from \cite{Guan2006} is slightly conservative under the null hypothesis, which can be attributed to the relatively small mean number of observed points in the simulated datasets and the test being asymptotic. Under alternative hypotheses, the test has power comparable to that of the \cite{SchlatherEtal2004} test.

\subsubsection*{Overview study: P-C tests}

Regarding the test of independence between the points and the covariate (P-C test), the \cite{SchlatherEtal2004} test is again slightly liberal under the null hypothesis (models M1, M2, M4, M6). The power of the test is higher for models M3 and M5 compared to models M7 and M8. This can be explained by the fact that in the former models, the covariate affects the point process directly, while in the latter models, the observed covariate is noisy, see Table~\ref{tab:models_overall}.

\citeauthor{Guan2006}'s test is slightly conservative under the null hypothesis while having almost the same power as \citeauthor{SchlatherEtal2004}'s test against the alternatives considered here.

As expected, the random shift with torus correction is liberal under the null hypothesis and has very high power in the rest of the cases. We remark that the liberality of the torus correction makes a direct comparison of power with other methods difficult. The random shift test with variance correction is slightly conservative under the null hypothesis and has slightly lower power than the test with torus correction, but always higher power than both \citeauthor{SchlatherEtal2004} and \citeauthor{Guan2006}'s tests. This is caused by the fact that the random shift tests take advantage of the knowledge of covariate values in the whole observation window, while \citeauthor{SchlatherEtal2004} and \citeauthor{Guan2006}'s tests use only the covariate values sampled at the points of the process.

\subsubsection*{Overview study: PM-C tests}

Concerning the test of independence between the marks and the covariate (PM-C test), the random shift tests with torus correction are all slightly liberal under the null hypo\-thesis (models M1, M3, M4, M5), regardless of the test statistic used, see Table~\ref{tab:results_overall}.

Using the variance correction reduces the liberality, but does not make the tests conservative. Note that \cite{MrkvickaEtAl2020} reported conservativeness of the random shift test with variance correction in independence tests between pairs of random fields. However, in \cite{MrkvickaEtAl2020}, the random fields were sampled at the points of a homogeneous binomial point process, while in the present paper the mark and covariate values are sampled at the points of a cluster point process, which brings bias into the estimation of their covariance structure, as reported by \cite{Diggle2010}.

It is important to note that the liberality is more severe for M3, M4, and M5 than for model M1. In the latter model, the points, marks, and the covariate are independent, while in models M3 to M5 the points are not independent of marks and the covariate, increasing the liberality of the tests. One can remedy this issue by using correlation coefficients instead of sample covariance as the test statistic. This is investigated in more detail in Sections~\ref{sec:preferentialSampling} and \ref{sec:dependentMarking}.

The performance of all PM-C tests under different alternatives is comparable (see Table~\ref{tab:results_overall}). The variance correction performs only marginally better than the torus correction, and the Pearson's correlation coefficient results in a slightly higher power than the other two test statistics.

\subsection{Influence of preferential sampling}\label{sec:preferentialSampling}

Here, we focus on testing independence of marks and a covariate under preferential sampling of the covariate values. Models M9 and M10 described in Table~\ref{tab:models_detailed} depend on a parameter $\alpha$ determining the degree of preferential sampling. The value $\alpha=0$ corresponds to the independence between the points and the covariate. Increasing values of $\alpha$ correspond to increasing dependence. The values of $\alpha$ are chosen to be 0.0, 0.2, 0.4, 0.6, 0.8, and 1.0, respectively. Note that the point process model M10 is no longer a log-Gaussian Cox process, but it is standardized to have the same intensity $\exp\{5\}$ as the previous models M1 to M8 and M9.

Apart from reporting the fractions of rejections for the PM-C tests, we also include in the tables of results the P-C and P-M tests so that the strength of dependence between the points and the covariate or the points and the marks in a given model can be assessed. For the P-C test, we use for clarity only the random shift approach with variance correction -- in the overall study it was not liberal, as opposed to the torus correction, and it had much higher power than \citeauthor{SchlatherEtal2004} and \citeauthor{Guan2006}'s tests. For the P-M test we use only \citeauthor{Guan2006}'s test -- in the overall study it was not liberal and had similar power to the \citeauthor{SchlatherEtal2004}'s test.

\begin{table}
\renewcommand{\arraystretch}{1.3}
\footnotesize
\begin{tabular}{|l|l|c|c|c|}
 \hline
  & Structure & Points & Marks & Covariate \\ 
 \hline\hline
   M9 & P--M--C+P & $\exp\{4.5 + \alpha Z_3(u) + \sqrt{1-\alpha^2} Z_1(u)\}$ & $Z_2(u)$ & $Z_3(u)$  \\ \hline
   M10 & P--M--C+P & $\exp\{-\alpha Z_3(u)^2+4.5+Z_1(u)+\log(1+2\alpha)/2\}$ & $Z_2(u)^2$ & $Z_3(u)^2$  \\ \hline \hline
  M11 & P+M--C--P & $\exp\{4.5 + \alpha Z_2(u) + \sqrt{1-\alpha^2} Z_1(u)\}$ & $Z_2(u)$ & $Z_3(u)$  \\ \hline
  M12 & P+M--C--P & $\exp\{-\alpha Z_2(u)^2+4.5+Z_1(u)+\log(1+2\alpha)/2\}$ & $Z_2(u)^2$ & $Z_3(u)^2$  \\ \hline  
\end{tabular}
\caption{Models for the detailed study of the influence of preferential sampling (top part) and dependent marking (bottom part). For clarity, in the se\-cond column the true dependence structure of the point locations (P), marks (M), and the covariate (C) is given for each model. The ``+'' sign indicates dependence between the given pair of objects, the ``-'' sign indicates independence.}
\label{tab:models_detailed}
\end{table}

\begin{table}
\small
\renewcommand{\arraystretch}{1.2}
\begin{tabular}{|l|c|c|c|c|c|c|}
 \hline
 $\alpha$ & 0.0 & 0.2 & 0.4 & 0.6 & 0.8 & 1.0 \\ \hline\hline
 P-M (Guan)           & 0.033 & 0.032 & 0.037 & 0.035 & 0.039 & 0.034 \\ \hline \hline 
 P-C (variance)       & 0.044 & 0.134 & 0.419 & 0.780 & 0.973 & 1.000 \\ \hline\hline
 PM-C (torus, cov)    & 0.071 & 0.072 & 0.071 & 0.069 & 0.074 & 0.074 \\ \hline
 PM-C (variance, cov) & 0.065 & 0.067 & 0.065 & 0.064 & 0.069 & 0.070 \\ \hline
 PM-C (torus, Pea)    & 0.072 & 0.073 & 0.073 & 0.067 & 0.078 & 0.081 \\ \hline
 PM-C (variance, Pea) & 0.053 & 0.058 & 0.057 & {\bf 0.052} & 0.061 & 0.067 \\ \hline
 PM-C (torus, Ken)    & 0.068 & 0.072 & 0.070 & 0.070 & 0.075 & 0.073 \\ \hline
 PM-C (variance, Ken) & {\bf 0.049} & {\bf 0.055} & {\bf 0.055} & 0.054 & {\bf 0.060} & {\bf 0.062} \\ \hline
\end{tabular}
\caption{Results of the detailed study for model M9 with different choices of $\alpha$: fraction of rejections of the corresponding null hypothesis. Results based on $5\,000$ independent replications. The confidence interval for the rejection rate based on the binomial distribution for the nominal significance level $0.05$ is [0.044,0.056].}
\label{tab:results_M9}
\end{table}

Model M9 is, in fact, a parametrized version of model M3. Marks are independent of the covariate. With $\alpha>0$, the points of the process occur more often at locations with high covariate values. In this case, the preferential sampling occurring for $\alpha > 0$ introduces a very mild liberality to the PM-C tests, growing only very slightly with an increasing value of $\alpha$, see Table~\ref{tab:results_M9}, showing the fraction of rejections of individual tests for model M9. We note that the actual significance level of the tests is close to the nominal 0.05 level. Tests with variance correction exhibit less liberality than the torus version, and similarly, tests based on sample correlation coefficients exhibit less liberality than tests based on sample covariance.

\begin{table}
\renewcommand{\arraystretch}{1.2}
\small
\begin{tabular}{|l|c|c|c|c|c|c|}
 \hline
 $\alpha$ & 0.0 & 0.2 & 0.4 & 0.6 & 0.8 & 1.0 \\ \hline\hline
 P-M (Guan)           & 0.061 & 0.059 & 0.055 & 0.062 & 0.058 & 0.059 \\ \hline \hline
 P-C (variance)       & 0.040 & 0.306 & 0.739 & 0.944 & 0.990 & 0.999 \\ \hline\hline
 PM-C (torus, cov)    & 0.059 & 0.010 & 0.003 & 0.000 & 0.000 & 0.000 \\ \hline
 PM-C (variance, cov) & 0.061 & 0.012 & 0.005 & 0.001 & 0.001 & 0.000 \\ \hline
 PM-C (torus, Pea)    & {\bf 0.056} & 0.035 & 0.037 & 0.024 & 0.025 & 0.023 \\ \hline
 PM-C (variance, Pea) & 0.059 & {\bf 0.039} & 0.043 & {\bf 0.030} & 0.029 & {\bf 0.027} \\ \hline
 PM-C (torus, Ken)    & 0.058 & 0.037 & 0.038 & 0.027 & 0.031 & 0.024 \\ \hline
 PM-C (variance, Ken) & 0.060 & {\bf 0.039} & {\bf 0.044} & {\bf 0.030} & {\bf 0.035} & 0.024 \\ \hline
\end{tabular}
\caption{Results of the detailed study for model M10 with different choices of $\alpha$: fraction of rejections of the corresponding null hypothesis. Results based on $5\,000$ independent replications. The confidence interval for the rejection rate based on the binomial distribution for the nominal significance level $0.05$ is [0.044,0.056].}
\label{tab:results_M10}
\end{table}

In model M10 the marks are again independent of the covariate. With $\alpha>0$, the points of the process occur more often at locations with low covariate values. Since the covariate corresponds to a squared Gaussian random field, its values are bounded from below by zero. Hence, the preferentially sampled covariate values are likely to be close to zero and the variance of the sample covariance is smaller than under non-preferential sampling. Since the random shifts produce data with behavior similar to non-preferential sampling, the tests based on the sample covariance are conservative (the observed value of test statistic $T_0$ is too rarely found to be extreme among the Monte Carlo values $T_1, \ldots, T_N$). This is clearly observed in Table~\ref{tab:results_M10}, showing the fraction of rejections of the individual tests for model M10, where the tests based on the sample covariance are more conservative with increasing value of $\alpha$. The other two test statistics (Pearson's and Kendall's correlation coefficients) are less affected by the variance of the sampled covariate values and hence are much less conservative. The performance of the tests based on Pearson's and Kendall's correlation coefficients is comparable. Interestingly, the variance correction is always less conservative than the torus correction for $\alpha > 0$.

\subsection{Influence of dependent marking}\label{sec:dependentMarking}

Here, we focus on testing the independence of marks and a covariate under dependent marking. Models M11 and M12 described in Table~\ref{tab:models_detailed} depend on a parameter $\alpha$ determining the strength of dependence between the points and the marks. The value $\alpha=0$ corresponds to the geostatistical marking model and the independence between the points and the covariate. Increasing values of $\alpha$ correspond to increasing dependence. The values of $\alpha$ are chosen to be 0.0, 0.2, 0.4, 0.6, 0.8, and 1.0, respectively. Note that the point process model M12 is not a log-Gaussian Cox process anymore, but it is standardized to have the same intensity $\exp\{5\}$ as the previous models M1 to M8 and M9, M11.
As in the previous section, we also report the results of the P-C test with variance correction and \citeauthor{Guan2006}'s P-M test so that the strength of dependence between the points and the covariate or between the points and the marks in a given model can be assessed.

\begin{table}
\small
\renewcommand{\arraystretch}{1.2}
\begin{tabular}{|l|c|c|c|c|c|c|}
 \hline
 $\alpha$ & 0.0 & 0.2 & 0.4 & 0.6 & 0.8 & 1.0 \\ \hline\hline
 P-M (Guan)           & 0.038 & 0.065 & 0.151 & 0.342 & 0.622 & 0.907 \\ \hline \hline
 P-C (variance)       & 0.043 & 0.036 & 0.041 & 0.040 & 0.043 & 0.038 \\ \hline\hline
 PM-C (torus, cov)    & 0.070 & 0.068 & 0.067 & 0.070 & 0.070 & 0.065 \\ \hline
 PM-C (variance, cov) & 0.060 & 0.064 & 0.064 & 0.065 & 0.070 & 0.065 \\ \hline
 PM-C (torus, Pea)    & 0.070 & 0.071 & 0.069 & 0.072 & 0.073 & 0.067 \\ \hline
 PM-C (variance, Pea) & 0.056 & {\bf 0.056} & 0.057 & 0.058 & {\bf 0.057} & 0.053 \\ \hline
 PM-C (torus, Ken)    & 0.071 & 0.071 & 0.066 & 0.073 & 0.071 & 0.064 \\ \hline
 PM-C (variance, Ken) & {\bf 0.051} & {\bf 0.056} & {\bf 0.051} & {\bf 0.057} & 0.058 & {\bf 0.051} \\ \hline
\end{tabular}
\caption{Results of the detailed study for model M11 with different choices of $\alpha$: fraction of rejections of the corresponding null hypothesis. Results based on $5\,000$ independent replications. The confidence interval for the rejection rate based on the binomial distribution for the nominal significance level $0.05$ is [0.044,0.056].}
\label{tab:results_M11}
\end{table}

\begin{table}
\small
\renewcommand{\arraystretch}{1.2}
\begin{tabular}{|l|c|c|c|c|c|c|}
 \hline
 $\alpha$ & 0.0 & 0.2 & 0.4 & 0.6 & 0.8 & 1.0 \\ \hline\hline
 P-M (Guan)           & 0.061 & 0.026 & 0.020 & 0.022 & 0.022 & 0.022 \\ \hline \hline
 P-C (variance)       & 0.046 & 0.042 & 0.047 & 0.048 & 0.049 & 0.049 \\ \hline\hline
 PM-C (torus, cov)    & 0.057 & {\bf 0.053} & 0.055 & 0.055 & 0.047 & 0.055 \\ \hline
 PM-C (variance, cov) & 0.064 & 0.060 & 0.060 & 0.059 & 0.047 & 0.053 \\ \hline
 PM-C (torus, Pea)    & {\bf 0.055} & 0.056 & 0.056 & 0.056 & 0.048 & 0.055 \\ \hline
 PM-C (variance, Pea) & 0.059 & 0.060 & 0.055 & 0.057 & 0.053 & 0.052 \\ \hline
 PM-C (torus, Ken)    & {\bf 0.055} & 0.054 & 0.054 & 0.053 & 0.048 & 0.053 \\ \hline
 PM-C (variance, Ken) & 0.060 & {\bf 0.053} & {\bf 0.052} & {\bf 0.052} & {\bf 0.050} & {\bf 0.050} \\ \hline
\end{tabular}
\caption{Results of the detailed study for model M12 with different choices of $\alpha$: fraction of rejections of the corresponding null hypothesis. Results based on $5\,000$ independent replications. The confidence interval for the rejection rate based on the binomial distribution for the nominal significance level $0.05$ is [0.044,0.056].}
\label{tab:results_M12}
\end{table}

Model M11 is, in fact, a parametrized version of model M4. Marks are independent of the covariate. With $\alpha>0$, the points of the process occur more often at locations with high values of the mark field. In this case, the dependent marking occurring for $\alpha > 0$ does not introduce further liberality to the PM-C tests compared to the case with $\alpha = 0$, see Table~\ref{tab:results_M11}, showing the fraction of rejections of the individual tests for model M11. The tests with variance correction exhibit less liberality than the torus version. In the tests using variance correction, the tests based on the sample correlation coefficients exhibit less liberality than the tests based on the sample covariance.

In model M12 the marks are again independent of the covariate. With $\alpha>0$, the points of the process occur more often at locations with low values of the mark field. For this model, the performance of the PM-C tests with the torus correction and the variance correction is comparable, and one is not uniformly better than the other; see Table~\ref{tab:results_M12}, showing the fraction of rejections of the individual tests for model M12. The same applies to tests based on the sample covariance and based on sample correlation coefficients. All tests match the nominal 0.05 significance level very closely, and their performance is satisfactory. Note that in this model, \citeauthor{Guan2006}'s P-M test cannot recognize the dependence between the points and the marks.

Note that while model M12 is a counterpart of model M10, we do not observe the conservativeness of the PM-C tests in this case. The difference is that in model M10 the dependence of the points and the covariate is broken by the random shifts, and new covariate values are sampled for each shift vector. This results in the variance of the test statistic computed from the original data being smaller than the variance of the test statistic computed from the shifted data; see the discussion in Section~\ref{sec:preferentialSampling}. This does not apply to model M12 where points and marks are shifted together, and no new mark values are sampled for each shift vector.

\subsection{Multiple covariates}\label{sec:multiple}

In this section, we study the possibility to test independence between marks and multiple covariates simultaneously. We consider a pair of covariates and a model structure similar to model M2 from Table~\ref{tab:models_overall}, where the points are independent of the marks and the covariates, to focus on the dependence between the marks and the covariates.

\begin{table}
\renewcommand{\arraystretch}{1.3}
\footnotesize
\begin{tabular}{|l|l|c|c|c|c|}
 \hline
  & Structure & Points & Marks & Covariate A & Covariate B \\ 
 \hline\hline
   M2.1 & P--M--C--P & $\exp\{4.5 + Z_1(u)\}$ & $a Z_2(u) + a Z_3(u)$ & $a Z_4(u) + a Z_6(u)$ & $a Z_5(u) + a Z_6(u)$  \\ \hline
   M2.2 & P--M+C--P  & $\exp\{4.5 + Z_1(u)\}$ & $a Z_2(u) + a Z_3(u)$ & $a Z_3(u) + a Z_6(u)$ & $a Z_4(u) + a Z_6(u)$  \\ \hline
   M2.3 & P--M+C--P  & $\exp\{4.5 + Z_1(u)\}$ & $a Z_2(u) + a Z_3(u)$ & $a Z_2(u) + a Z_6(u)$ & $a Z_3(u) + a Z_6(u)$  \\ \hline
\end{tabular}
\caption{Models for the study of independence testing between marks and a pair of covariates. The considered models are variants of model M2 from Table~\ref{tab:models_overall}. For clarity, in the se\-cond column the true dependence structure of the point locations (P), marks (M), and the covariate (C) is given for each model. The ``+'' sign indicates dependence between the given pair of objects, the ``-'' sign indicates independence.}
\label{tab:models_detailed_multiple}
\end{table}

Models M2.1 to M2.3 are described in Table~\ref{tab:models_detailed_multiple}. Assuming that $Z_1, \ldots, Z_6$ are independent Gaussian random fields with the distribution given at the beginning of Section~\ref{sec:simulations}, marks are influenced by $Z_2$ and $Z_3$ in all three models. Covariates A and B are either constructed from independent random fields (M2.1), or one of them is constructed from $Z_3$ and the other one from an independent random field (M2.2), or one covariate is constructed from $Z_2$ and the other from $Z_3$. In all cases, both covariates contain noise described by the same random field $Z_6$ to introduce dependence between the covariates. Model M2.2 corresponds directly to the original model M2, only with the extra covariate not influencing the marks. Similarly, model M2.1 corresponds to the original model M1.

The tests of independence between the marks and the pair of covariates are based on random shifts. For each shift, a bivariate test statistic value is computed according to formulas in Section~\ref{subsec:PMC}, where the elements of the bivariate vector correspond to the individual covariates. The outcome of the test can be determined from the bivariate test statistic values using the global envelope test \citep{MyllymakiEtal2017}. Alternatively, two tests for individual covariates may be performed separately and then linked through the Bonferroni correction. Both of these approaches are considered here.

The results are given in Table~\ref{tab:results_M2_multiple}. We observe that both types of multiple testing correction (Bonferroni correction and the global envelope test) perform equally well, despite the dependence between the two covariates. For model M2.1, with independence between the marks and the pair of covariates, the results indicate slight liberality of the torus correction which is reduced by using the variance correction and either Pearson's or Kendall's correlation coefficient. This is consistent with the results for model M1 in Table~\ref{tab:results_overall}, also given in the second column of Table~\ref{tab:models_detailed_multiple} for easier comparison. If one or both covariates influence the marks (models M2.2 and M2.3, respectively), the tests are very successful at recognizing the dependence, even though only noisy observations are available, with the tests based on the Pearson's correlation coefficient achieving the highest power.

Model M2.2 can be directly compared with the original model M2 in order to assess the effect of the extra covariate not influencing the marks. Table~\ref{tab:results_M2_multiple} contains an additional column (the fifth column) containing the corresponding results for model M2, taken from Table~\ref{tab:results_overall}, to facilitate comparison. As expected, we observe slight loss of power when including the extra covariate; however, the loss is not very prominent.

\begin{table}
\small
\renewcommand{\arraystretch}{1.2}
\begin{tabular}{|l||c||c|c||c||c|c|c|c|}
 \hline
  & M1 & \multicolumn{2}{c||}{M2.1} & M2 & \multicolumn{2}{c|}{M2.2} & \multicolumn{2}{c|}{M2.3} \\ \hline
  &  & Bon & GET &  & Bon & GET & Bon & GET \\ \hline\hline
 PM-C (torus, cov)    & 0.061 & 0.067 & 0.069 & 0.785 & 0.723 & 0.724 & 0.872 & 0.875 \\ \hline
 PM-C (variance, cov) & 0.059 & 0.056 & 0.059 & 0.799 & 0.736 & 0.735 & 0.877 & 0.878 \\ \hline
 PM-C (torus, Pea)    & 0.067 & 0.065 & 0.067 & 0.826 & 0.782 & 0.782 & 0.913 & 0.913 \\ \hline
 PM-C (variance, Pea) & 0.053 & {\bf 0.045} & {\bf 0.047} & {\bf 0.835} & {\bf 0.795} & {\bf 0.795} & {\bf 0.921} & {\bf 0.921} \\ \hline
 PM-C (torus, Ken)    & 0.062 & 0.062 & 0.065 & 0.794 & 0.745 & 0.745 & 0.893 & 0.894 \\ \hline
 PM-C (variance, Ken) & {\bf 0.050} & 0.044 & 0.046 & 0.804 & 0.743 & 0.742 & 0.891 & 0.891 \\ \hline
\end{tabular}
\caption{Results of the study with multiple covariates: fraction of rejections of the null hypothesis of independence between marks and the pair of covariates. Results based on $5\,000$ independent replications. The confidence interval for the rejection rate based on the binomial distribution for a nominal significance level $0.05$ is [0.044,0.056].}
\label{tab:results_M2_multiple}
\end{table}

\subsection{Implementation and software}\label{subsec:software}

When implementing the \cite{SchlatherEtal2004} test, we take advantage of the \texttt{R} package \texttt{spatstat} \citep{baddeley2015spatialR} and its function \texttt{Emark}, which computes the estimate $\widehat{E}(t)$, and the function \texttt{orderNorm} from the \texttt{bestNormalize} package, which performs the marginal transformation of the observed mark values to a normal distribution \citep{bestNormalizePackage}. The package \texttt{geoR} \citep{DiggleRibeiro2007} is used to estimate the covariance structure of the marks (or covariate, respectively) in both \citeauthor{SchlatherEtal2004} and \citeauthor{Guan2006}'s tests. The latter test uses a modification of the \texttt{Kinhom} function from the \texttt{spatstat} package to estimate the required quantities. The random shift methods use our implementation, the source codes being available for download at \url{http://msekce.karlin.mff.cuni.cz/~dvorak/software.html}.

\section{Applications}\label{sec:applications}

\subsection{Castilla-La Mancha forest fires}\label{sec:clm}

We illustrate the proposed procedure on a real dataset consisting of locations $(x_i)$ and size measurements ($m_i$) of 689 forest fires that occurred in the Castilla-La Mancha region (Spain) in 2007. The size of the region is approximately 400 by 400 kilometers. The mark associated with a forest fire is the total area burned by the fire (in hectares). It is part of a larger dataset available in the \texttt{spatstat} package \citep{baddeley2015spatialR}.

The dataset is plotted in Figure~\ref{fig:clm_intro} (left) together with terrain elevation in the given region (right). A short preliminary inspection of the relationship between the elevation (covariate) and the total area burned in individual fires (marks) suggests that fires occurring at higher elevation tend to burn smaller areas. This is illustrated by the empirical Kendall's correlation coefficient between the mark values and the covariate values sampled at the data points being -0.144. This effect appears to be justified by the structure of vegetation changing with altitude. Below we perform formal tests of the hypothesis of independence between marks and the covariate, but we first look into possible preferential sampling and dependent marking to investigate the complete dependence structure in the dataset.

When performing different tests discussed in this paper, we follow the choices made in the simulation study in Section~\ref{sec:simulations}, i.e. the random shift tests are based on 999 shifts, with radius 150 kilometers for this dataset. The \cite{Guan2006} test is used with an upper bound of 40 meters. Note that due to the irregular shape of the observation window, random shift tests with torus correction are not applicable here. We do not perform here the \cite{SchlatherEtal2004} test -- it is difficult to specify an appropriate model for the covariance structure for the real dataset, and this is not the focus of our approach.

\subsubsection{P-C test}

We first investigate the possible preferential sampling of the covariate values, as discussed in Section~\ref{subsec:PC}. The random shift test (with variance correction) of independence between points and covariate, using the sample mean as the test statistic, results in a p-value of 0.684. Thus, we have no significant evidence against the null hypothesis. For completeness, \citeauthor{Guan2006}'s test results in a p-value of 0.614, confirming the previous conclusion.


\subsubsection*{P-M test}

We test the independence between the marks and points using \citeauthor{Guan2006}'s test. It results in a p-value of $8.7 \cdot 10^{-7}$, rejecting the null hypothesis of the geostatistical marking model, implying dependencies between point positions and the respective marks.



\subsubsection*{PM-C test}

The final step of the analysis is to test the independence between the marks and the covariate. The results obtained so far indicate that the dataset at hand follows the structure either of model M4 or M6 from the simulation study, see Table~\ref{tab:models_overall}. The results of the si\-mu\-la\-tion study in Table~\ref{tab:results_overall} suggest that using the random shift test with variance correction and Kendall's correlation coefficient as the test statistic is the most appropriate, the test exhibiting almost no liberality. The test results in a p-value of 0.108, providing no significant evidence against the null hypothesis. Other tests considered in the simulation study in Section~\ref{sec:simulations} confirm the same conclusion -- the p-value of the random shift test with variance correction and sample covariance (Pearson's correlation coefficient) was 0.580 (0.980).

If the analysis aimed to investigate the overall dependence structure of the points, marks, and the covariate, we would choose in advance the three tests (random shift test with variance correction for the P-C test, \cite{Guan2006} test for P-M, random shift with variance correction and Kendall's correlation coefficient for PM-C) and bind the tests together e.g. by the Bonferroni correction. With this correction, one of the three tests yields significant results at the 5\% significance level (the P-M test) and we reject the null hypothesis of independence between points, marks, and the covariate.

\subsection{Crimes in Kennedy-Bogota}\label{sec:bogota}
Now we provide an example of a possible adaptation of the methods proposed in this paper to point patterns with categorical marks. Kennedy, the eighth neighborhood of Colombia's capital, is located in the southwest of the city and its population is approximately $1\,200\,000$ inhabitants (www.bogota.gov.co). Kennedy is particularly known as one of the most problematic neighborhoods from a social point of view. It reports a high degree of poverty and public disorder, as well as a high crime record. Data were collected by the National Police as part of the administrative routines held by the agents when crimes are investigated.

\begin{figure}[tb]
    \centering
    \includegraphics[width=0.6\textwidth]{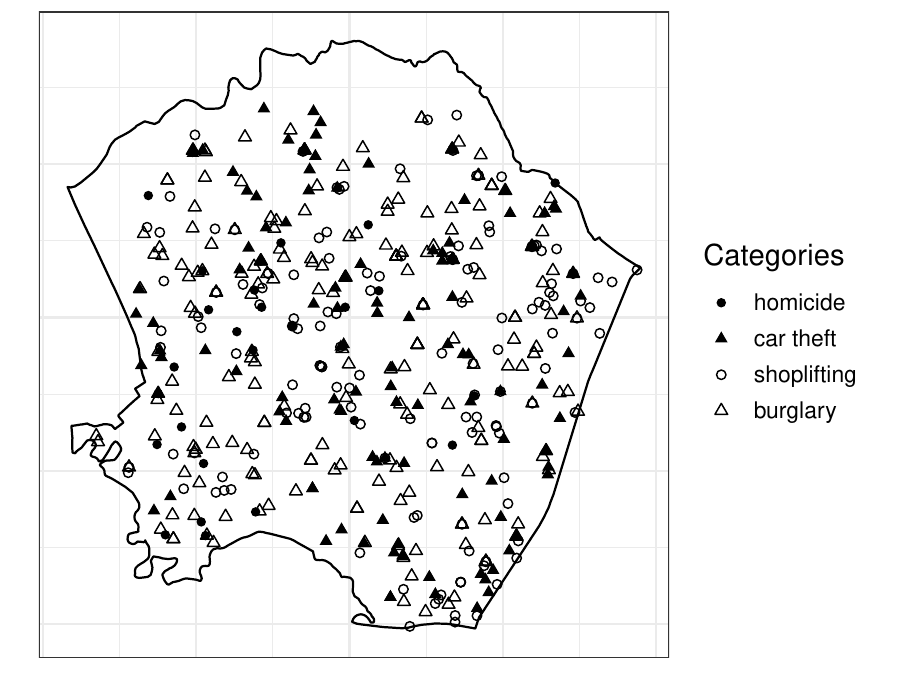}
    \caption{Bogota crime dataset (a random sample of 500 reports). Spatial locations and types reported in Kennedy from 2012 to 2017.}
    \label{fig:bogota_sample}
\end{figure}

Our data consist of the locations of $13\,959$ reported crimes in Kennedy between 2012 and 2017. The crimes are divided into four different categories which serve as marks: homicide, car theft, shoplifting, and burglary. A small random subsample of the point pattern is displayed in Figure \ref{fig:bogota_sample}.

\begin{figure}[tb]
    \centering
    \includegraphics[width=0.8\textwidth]{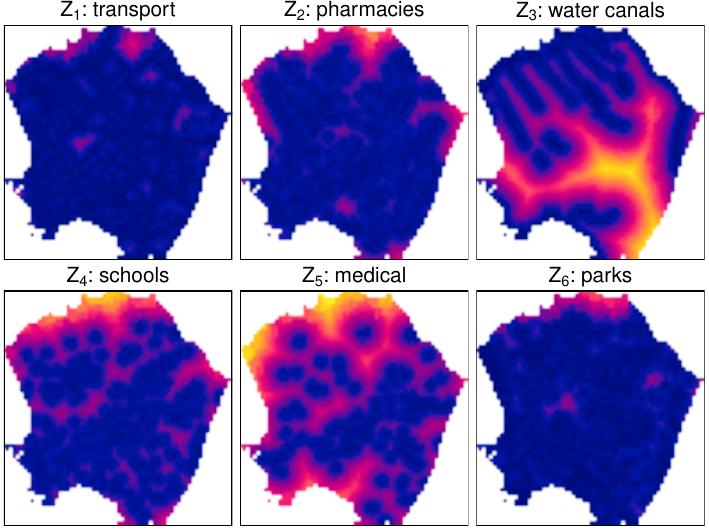}
    \caption{Bogota crime dataset. The six covariates considered in this study, from top to bottom, left to right: distance to the nearest transport service, pharmacy, water canal, school, health service and park, respectively. The color scale is the same for all the plots, ranging from 0 (darkest color) to 1.55 km (brightest color).}
    \label{fig:bogota_covariates}
\end{figure}

In addition, there are six covariates in our study. They are defined in terms of distances to fixed locations, namely the nearest transport service, nearest pharmacy, nearest water canal, nearest school, nearest medical centre, and nearest park. The continuous covariate maps were obtained by spatial smoothing of discrete measurements using a Nadaraya-Watson approach (see, e.g. \citealp{baddeley2015spatialR} and references therein) with a bandwidth chosen by least-squares cross-validation \citep[p.49]{silverman1986density}. The lots of the covariates are shown in Figure~\ref{fig:bogota_covariates}. All random shift tests are performed with the variance correction, since the torus correction is not available due to the irregular observation window. We use shift vectors generated uniformly on a disc with a radius of 3.5 km (the width and height of the observation window is approximately 7.5 km) and 999 independent random shifts.
Again, we do not perform here the \cite{SchlatherEtal2004} test, it is difficult to specify an appropriate model for the covariance structure for the real dataset.

\subsubsection*{P-M test}

In testing independence between the point locations and categorical marks, it is not natural to formulate the null hypothesis in terms of the geostatistical marking model. Instead, random labeling or random superposition hypothesis is more appropriate \citep{Diggle2010a}. For the current dataset, we consider the latter hypothesis, stating that the marked point process is a superposition of independent component processes.

The random superposition hypothesis in a bivariate point process can be tested using the random shift approach with the kernel method of variance correction  \citep[see ][Sec. 2.3]{MrkvickaEtAl2020}. We performed the test for each pair of component processes corresponding to different mark values. As suggested in the paper, we use the expected distance to the nearest neighbor with a different mark as the test statistic. No pair of component processes is found to be dependent on the 5~\% significance level, the smallest p-value being 0.084 for the test of independence between car thefts and burglaries. Consequently, the null hypothesis of random superposition is not rejected.

\subsubsection*{P-C test}

When testing independence between point locations (disregarding the marks) and covariates, we use the random shift test with variance correction. When testing each covariate separately, we obtain the p-values 0.928, 0.032, 0.876, 0.862, 0.758 and 0.560, respectively, for covariates $Z_1$ to $Z_6$. On a 5\% significance level, only the covariate $Z_2$ (distance to the nearest pharmacy) is found significant. When testing independence between the points and all the six covariates simultaneously by the global envelope test, we obtain a p-value of 0.146, i.e. no covariate is found significant. This conclusion corresponds to the Bonferroni correction applied to the set of individual tests.

\begin{figure}[tb]
    \centering
    \includegraphics[trim={0.5cm 2cm 0 0.5cm},clip, width=0.8\textwidth]{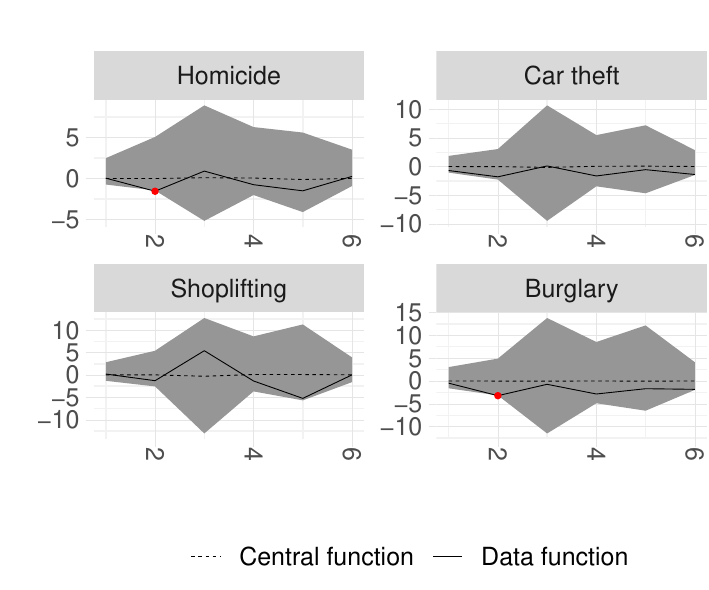}
    \caption{The crime dataset, output of the global envelope test of independence between points in individual component processes (separated according to the mark values) and all the six covariates considered in this study. The solid line gives the data values, the dashed line gives the mean functions. The horizontal axis corresponds to the six covariates.}
    \label{fig:bogota_PC_subpatterns}
\end{figure}

To gain a deeper understanding, we perform the test of independence between point locations and the individual covariates in the four component processes obtained by splitting the original dataset according to the mark values. These 24 tests, one for each combination of a component process and a covariate, can be bound together by a global envelope test procedure; see Figure~\ref{fig:bogota_PC_subpatterns}, obtaining a p-value of 0.034. The covariate $Z_2$ (distance to the nearest pharmacy) is significant for the processes of homicides and burglaries.
This test cannot be considered as a test of independence between marks and a set of covariates, since it considers in each of the 24 subtests only points of one type and a covariate. On the other hand, the test described in the next subsection considers in each subtest points of two types and a covariate. Nevertheless, the above test can be of value when one wants to understand the structure of the marginal point patterns.

For comparison, we perform \citeauthor{Guan2006}'s test for the individual covariates and the unmarked point pattern with an upper bound of 750 m. The resulting p-values are 0.475, $5 \cdot 10^{-4}$, 0.233, $4 \cdot 10^{-3}$, $3 \cdot 10^{-3}$ and 0.172, respectively, for covariates $Z_1$ to $Z_6$. On the 5\% significance level, the covariates $Z_2, Z_4$ and $Z_5$ are significant, even with the Bonferroni correction. The covariate $Z_2$, which is found to be significant by the random shift test, has the smallest p-value. The covariates $Z_4$ (distance to the nearest school) and $Z_5$ (distance to the nearest medical service) are not significant according to the random shift test.
In case of these two covariates, the random shift approach seems to be more appropriate because stationarity of only one of the objects (point process, covariate) is required; however, \citeauthor{Guan2006}'s test requires stationarity of both of them.


\subsubsection*{PM-C test}

The test of independence between the marks (the type of crime) and a covariate can be performed as suggested in Section~\ref{subsec:PMC}. The variance correction described in Section~\ref{sec:variance_correction} is applied to the individual means, and the differences of the standardized means are taken as the test statistic, as suggested in Section~\ref{subsec:PMC}.

When testing each of the six covariates individually, we obtain the p-values 0.278, 0.011, 0.078, 0.785, 0.027 and 0.041, respectively, for covariates $Z_1$ to $Z_6$. At the 5\% significance level, the covariates $Z_2, Z_5$ and $Z_6$ (distance to the nearest pharmacy, medical service and park, respectively) are found to be significant. With Bonferroni correction for multiple testing, no covariate is significant.

\begin{figure}[tbp]
    \centering
    \includegraphics[trim={0.5cm 2cm 0 0.5cm},clip, width=0.8\textwidth]{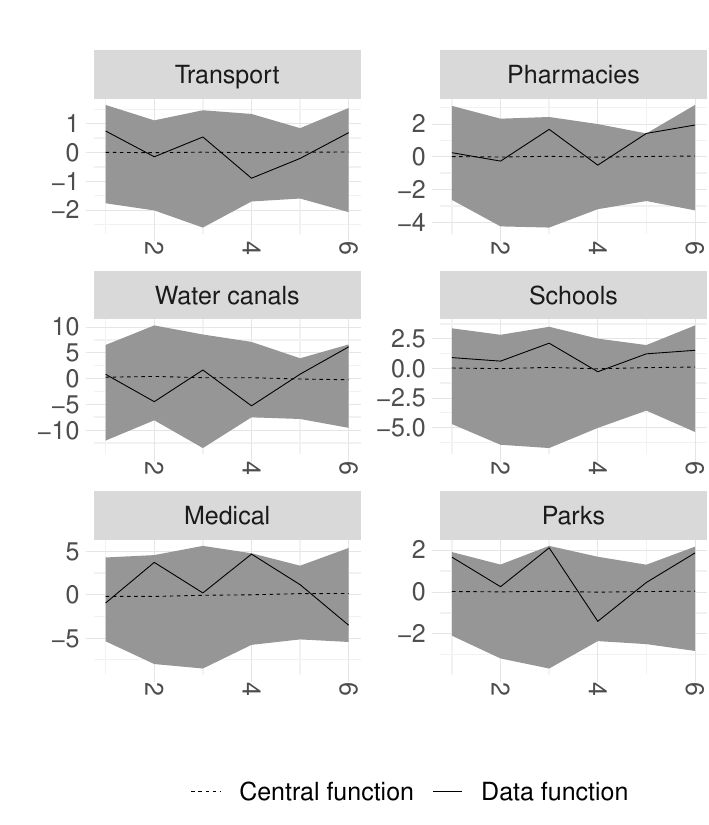}
    \caption{Bogota crime dataset. The output of the global envelope test of independence between categorical marks and all the six covariates simultaneously. No significant difference between the means of any pair of mark levels is found for any of the covariates. The solid line gives the data values, the dashed line gives the mean functions. The horizontal axis corresponds to the 6 differences between 4 types of crimes.}
    \label{fig:bogota_PMC_simultaneous}
\end{figure}

When testing the independence between the marks and the six covariates simultaneously using the global envelope test, we obtain a p-value of 0.051. This is in agreement with the tests of the individual covariates, with Bonferroni correction, reported in the previous paragraph. See Figure~\ref{fig:bogota_PMC_simultaneous} for the output of the test.

\section{Conclusions and discussion}\label{sec:conclusions}

In this paper, we investigate the dependence between the marks of a point process and a covariate. We have found that the problem involves the interaction of all three components, i.e. points, marks, and covariate, and that it is important to consider the dependence of all three possible pairs at once. For points and marks, the \cite{Guan2006} test seems to be the most appropriate. We have found that the random shift approach with variance correction is the most appropriate for points and a covariate. It has higher power than other methods considered in this study, since it also uses the information about covariate values outside the points, while not being liberal as the torus correction.

For marks and a covariate, the most appropriate method seems to be the random shift approach with the variance correction with Kendall's correlation coefficient as the test statistic, since it is more robust with respect to possible preferential sampling than other variants. It is also more exact than the random shift with the torus correction due to the absence of artificial modifications in the correlation structure of the data. 

We remark that there is no typical way a preferential sampling 
affects the PM-C test -- it can cause either liberality (see model M9 in Table~\ref{tab:results_M9}) or conservativeness of the test (see model M10 in Table~\ref{tab:results_M10}). Our simulation study also suggests that the dependent marking does not bring strong liberality or conservativeness to the PM-C test (see models M11, M12 in Tables~\ref{tab:results_M11},~\ref{tab:results_M12}).

We see the main benefit of the tests proposed in this paper in their nonparametric nature and in providing information about which covariates are relevant for further steps of statistical analysis. The tests provide similar but not identical information compared to, say, fitting a nonhomogeneous Poisson process. Note that the proposed tests account for possible interactions between the points and avoid full model specification. Also, while it is possible to test the significance of several covariates at once, as illustrated in Section~\ref{sec:multiple}, such an approach does not allow one covariate to compensate the influence of another one, as is the case in a parametric model for a nonhomogeneous Poisson process containing several covariates.

The procedures proposed in this paper were discussed in detail for numeric (continuous) marks and a numeric covariate. Categorical marks and a numeric covariate were briefly treated in the data example in Section~\ref{sec:bogota}. For a categorical covariate, the methods can be easily adapted using an appropriate vector test statistic based e.g. on the means of mark values in different subregions delineated by the discrete covariate values (for numeric marks) or counts of points of different types in individual subregions (for categorical marks). In this case, the torus correction can be easily applied for rectangular observation windows, but the variance correction cannot be used in the form presented in Section~\ref{sec:variance_correction} because it is not clear what would be an appropriate correction factor. However, the very general kernel regression approach to estimating the variance described by \cite{MrkvickaEtAl2020} can be used to determine the correction factors, and the variance correction approach (in the kernel version) remains applicable.

Even though we assumed in this paper that both $\Psi$ and $Z$ are stationary, it is enough if only one of them is stationary so that the random shift strategies are legitimate -- the stationary object is shifted while the non-stationary one is kept fixed.

The presented simulation experiments suggest that Pearson or Kendall's correlation coefficients are more robust with respect to possible dependence between points and marks or covariate than the sample covariance. Therefore, we recommend using Pearson or Kendall's correlation coefficient when the independence of two random fields is tested, even though \cite{MrkvickaEtAl2020} considered the sample covariance only. The preferential sampling issues may appear in the geostatistical problems, and \cite{MrkvickaEtAl2020} did not consider the robustness with respect to preferential sampling in their study.

\section*{Acknowledgements} This work was supported by Grant Agency of the Czech Republic (Project No.\ 19-04412S). J. Mateu and J.~A.~Gonz\'alez were partially supported by grants MTM2016-78917-R, AICO/2019/198, and UJI-B2018-04.

\bibliography{Tomas_Bibfile}
\bibliographystyle{agsm}

\appendix

\section{Variance correction factors -- proofs}

\subsection{Proof of Theorem~\ref{T1}}\label{app:proof1}

We begin by considering the mean and variance of $S=\sum_{x \in \Phi \cap W} Z(x)$. Conditioning on the realization of $Z$ and using Campbell's Theorem, we obtain $\mathbb{E}[S|Z] = \int_W \lambda Z(u) \,\mathrm{d}u$, and then by Fubini's Theorem, we get
\begin{align*}
    \mathbb{E} S = \mathbb{E} \Big [ \mathbb{E}[S|Z] \Big ] = \mathbb{E} \int_W \lambda Z(u) \,\mathrm{d}u  = \int_W \lambda  \mathbb{E} Z(u) \,\mathrm{d}u = 0.
\end{align*}
The last equality holds because we assume $\mathbb{E} Z(u)=0, u \in W$. Similarly,
\begin{align*}
    \mathrm{var}(S) = \mathbb{E} S^2 & = \int_W \int_W \lambda^2 g(u-v) \mathbb{E} Z(u)Z(v) \,\mathrm{d}u\,\mathrm{d}v + \int_W \mathbb{E} \lambda Z(u)^2 \,\mathrm{d}u \\
    & = \int_W \int_W \lambda^2 g(u-v) C(u-v) \,\mathrm{d}u\,\mathrm{d}v + \lambda |W| C(0).
\end{align*}

The upper bound for $\mathrm{var}(S)$ can be obtained using the assumption of bounded values ($C(u) \leq C(0)$ by the Cauchy-Schwarz inequality) and bounded support of $C$:
\begin{align*}
    \mathrm{var}(S) & \leq \int_W \int_W \lambda^2 C(0) g(u-v) \mathbb{I}_{B(o,R)}(u-v) \,\mathrm{d}u\,\mathrm{d}v + \lambda |W| C(0) \\
    & \leq \lambda^2 C(0) \int_W \int_{\mathbb{R}^2} g(u-v) \mathbb{I}_{B(o,R)}(u-v) \,\mathrm{d}u\,\mathrm{d}v + \lambda |W| C(0) \\
    & = \lambda^2 C(0) |W| K(R) + \lambda |W| C(0) = \lambda |W| \Big (\lambda C(0) K(R) + C(0) \Big ),
\end{align*}
where $\mathbb{I}$ is the indicator function, $B(o,R)$ is the open ball with radius $R$ centered at the origin $o$ and $K(R) = \int_{B(o,R)}g(u) \, \mathrm{d}u$. The lower bound of the same order can be obtained simply as $\mathrm{var}(S) \geq \lambda |W| C(0).$ Setting $C_1 = C(0)$ and $C_2 = \lambda C(0) K(R) + C(0)$ completes the proof of the first part of the theorem.

Considering now $U = \Phi(W)$, the approach of \citet[p.351]{StuartOrd1994}, based on the first-order Taylor expansion of the function $f(S,U) = S/U$, provides an approximation of the variance of the ratio $S/U$:
\begin{align*}
    \mathrm{var} \left( \frac{S}{U} \right) \approx \frac{1}{(\mathbb{E}U)^2} \mathrm{var}(S) - \frac{2 \mathbb{E} S}{(\mathbb{E}U)^3} \mathrm{cov}(S,U) + \frac{(\mathbb{E}S)^2}{(\mathbb{E}U)^4} \mathrm{var}(U).
\end{align*}
This simplifies greatly in our situation since $\mathbb{E}S=0$ and only the first term applies. Realizing that $\mathbb{E} U = \mathbb{E} \Phi(W) = \lambda |W|$ finishes the proof.

\subsection{Proof of Theorem~\ref{T2}}\label{app:proof2}

We begin by considering the mean and variance of
\begin{align*}
    S=\sum_{x,y \in \Phi \cap W}^{\neq}\mathrm{sgn}(Z_1(x) - Z_1(y)) \, \mathrm{sgn}(Z_2(x) - Z_2(y)).
\end{align*}
Conditioning on the realizations of $Z_1,Z_2$ and using Campbell's Theorem we obtain $\mathbb{E}[S|Z_1,Z_2] = \int_W \int_W \lambda_2(u,v) \mathrm{sgn}(Z_1(u) - Z_1(v)) \, \mathrm{sgn}(Z_2(u) - Z_2(v)) \,\mathrm{d}u \,\mathrm{d}v$ and then by Fubini's Theorem we get
\begin{align*}
    \mathbb{E} S & = \mathbb{E} \Big [ \mathbb{E}[S|Z_1,Z_2] \Big ] \\ 
    & = \mathbb{E} \int_W \int_W \lambda_2(u,v) \mathrm{sgn}(Z_1(u) - Z_1(v)) \mathrm{sgn}(Z_2(u) - Z_2(v)) \,\mathrm{d}u \,\mathrm{d}v \\
    & = \int_W \int_W \lambda_2(u,v) \mathbb{E} \Big [ \mathrm{sgn}(Z_1(u) - Z_1(v)) \mathrm{sgn}(Z_2(u) - Z_2(v)) \,\mathrm{d}u \,\mathrm{d}v \Big ]= 0.
\end{align*}
The last equality holds because the expected value is 0 due to the independence and symmetry of the distribution of $Z_1(u) - Z_2(v)$ and $Z_2(u) - Z_2(v)$.

We use similar arguments to determine $\mathrm{var}(S) = \mathbb{E}S^2$. To simplify notation in some parts of the proof, we denote 
\begin{align*}
    f(u,v)=\mathrm{sgn}(Z_1(u) - Z_1(v)) \, \mathrm{sgn}(Z_2(u) - Z_2(v)),
\end{align*}
noting that $f$ is a bounded symmetric function, i.e. $f(u,v)=f(v,u)$. By standard arguments, we obtain
\begin{align*}
    S^2 & = \left( \sum_{x,y \in \Phi \cap W}^{\neq} f(x,y) \right)^2 = I_1 + 4 I_2 + 2 I_3, \\
    I_1 & = \sum_{x,y,z,w \in \Phi \cap W}^{\neq} f(x,y) f(z,w) \\
    I_2 & = \sum_{x,y,z \in \Phi \cap W}^{\neq} f(x,y) f(x,z) \\
    I_3 & = \sum_{x,y \in \Phi \cap W}^{\neq} f(x,y)^2.
\end{align*}
Again using conditioning on the realizations of $Z_1,Z_2$, by Campbell's and Fubini's Theorems, we get
\begin{align*}
    \mathbb{E} I_1 & = \int_{W^4} \lambda_4(x,y,z,w) \mathbb{E} \Big [ f(x,y)f(z,w) \Big] \,\mathrm{d}x \,\mathrm{d}y \,\mathrm{d}z \,\mathrm{d}w, \\
    \mathbb{E} I_2 & = \int_{W^3} \lambda_3(x,y,z) \mathbb{E} \Big [ f(x,y)f(x,z) \Big] \,\mathrm{d}x \,\mathrm{d}y \,\mathrm{d}z, \\
    \mathbb{E} I_3 & = \int_{W^2} \lambda_2(x,y) \mathbb{E} \Big [ f(x,y)^2 \Big] \,\mathrm{d}x \,\mathrm{d}y.
\end{align*}

Now we investigate in detail the function $h(x,y,z,w) = \mathbb{E} f(x,y)f(z,w) $. Independence of $Z_1,Z_2$ implies that 
\begin{align*}
    h(x,y,z,w)=h_1(x,y,z,w)h_2(x,y,z,w)
\end{align*}
where
\begin{align*}
    h_1(x,y,z,w) & = \mathbb{E} \mathrm{sgn}(Z_1(x) - Z_1(y)) \, \mathrm{sgn}(Z_1(z) - Z_1(w)) \\
    h_2(x,y,z,w) & = \mathbb{E} \mathrm{sgn}(Z_2(x) - Z_2(y)) \, \mathrm{sgn}(Z_2(z) - Z_2(w)).
\end{align*}
We denote $X_1 = Z_1(x) - Z_1(y)$ and $X_2 = Z_1(z) - Z_1(w)$. The joint distribution of $\mathbb{X}=(X_1,X_2)^T$ is bivariate normal with the probability density function $f_{\mathbb{X}}$. We further denote by $E$ the union of the first and third quadrants in $\mathbb{R}^2$ and by $F$ the union of the second and fourth quadrant in $\mathbb{R}^2$. Then
\begin{align*}
    h_1(x,y,z,w) & = \mathbb{E} \mathrm{sgn}(X_1) \, \mathrm{sgn}(X_2) \\
    & = 1 \cdot \int_E f_{\mathbb{X}}(x_1,x_2) \,\mathrm{d}x_1 \,\mathrm{d}x_2 - 1 \cdot \int_F f_{\mathbb{X}}(x_1,x_2) \,\mathrm{d}x_1 \,\mathrm{d}x_2 \\
    &  \begin{cases}
= 0, \text{ if } \mathrm{cov}(X_1,X_2) = 0, \\
> 0, \text{ if } \mathrm{cov}(X_1,X_2) > 0, \\
< 0, \text{ if } \mathrm{cov}(X_1,X_2) < 0.
\end{cases}
\end{align*}
The same can be done for $h_2(x,y,z,w)$. From the assumption that $Z_1,Z_2$ have the same covariance function we get that the values $h_1(x,y,z,w)$ and $h_2(x,y,z,w)$ are either both 0, both positive or both negative. It follows that the function $h(x,y,z,w)$ is non-negative.

An upper bound for $\mathbb{E}I_1$ can be obtained from the assumption that $\lambda_4$ is bounded from above and from an upper bound on the volume of the set $V \subset W^4$ on which $h(x,y,z,w) > 0$. Recall that the covariance function $C$ has a bounded support, which is contained in the closed ball with radius $R$ centered in the origin, and note that
\begin{align}\label{eq:T2.covX}
    \mathrm{cov}(X_1,X_2) = C(x-z) + C(y-w) - C(y-z) - C(x-w).
\end{align}
Letting $x,y \in W$ be arbitrary, it is necessary (but not sufficient) for $\mathrm{cov}(X_1,X_2)$ to be non-zero that either $z \in W \cap \left( B(x,R) \cup B(y,R) \right)$ and $w \in W$ or vice versa. This implies that the volume of $V$ is at most $4\pi R^2 |W|^3$ and consequently there is a finite positive constant $M_1$ such that $\mathbb{E}I_1 \leq M_1 |W|^3$.

A lower bound for $\mathbb{E}I_1$ can be obtained from the assumption that $\lambda_4$ is bounded from below by a positive constant and from a lower bound on the volume of the set $V$. Such a lower bound can be found by considering $V_0 \subset V$ defined as
\begin{align*}
    V_0 = \big \{ (x,y,z,w) \in W^4: \; & x \in W, y \notin B(x,2R), z \in  B(x,r) \cup B(y,r), \\
    & w \notin  B(x,R) \cup B(y,R) \big \}
\end{align*}
For $(x,y,z,w) \in V_0$ only one term in \eqref{eq:T2.covX} is non-zero and $|\mathrm{cov}(X_1,X_2)| \geq \delta$, implying that $h(x,y,z,w) \geq \delta_0 > 0$ for some $\delta_0$ not depending on $(x,y,z,w)$. When $R$ is small compared to the size of W there is a constant $a > 0$ such that $|V_0| \geq a |W|^3$ and consequently there is a finite positive constant $m_1$ such that $\mathbb{E}I_1 \geq m_1 |W|^3$.

As above, we investigate the properties of $\mathbb{E} f(x,y)f(x,z)$ as a function of $(x,y,z) \in W^3$ and, as a consequence, the properties of $\mathbb{E}I_2$. In this case, the analogue of the function $h_1$ is $\mathbb{E} \mathrm{sgn}(Z_1(x) - Z_1(y)) \,  \mathrm{sgn}(Z_1(x) - Z_1(z))$ and similarly for $h_2$. Denoting $Y_1 = Z_1(x) - Z_1(y), Y_2 = Z_1(x) - Z_1(z)$, it holds that
\begin{align}\label{eq:T2.covY}
    \mathrm{cov}(Y_1,Y_2) = C(0) - C(x-y) - C(x-z) + C(y-z).
\end{align}
The key difference between covariances \eqref{eq:T2.covX} and \eqref{eq:T2.covY} is the term $C(0)$ in the latter. This causes the same order of $\mathbb{E}I_2$ and $\mathbb{E}I_1$, despite the different dimensionality of the integrals. Using the same arguments as above, we find finite positive constants $m_2, M_2$ such that $m_2 |W|^3 \leq \mathbb{E}I_2 \leq M_2 |W|^3$.

Considering $\mathbb{E}I_3$, from the assumption of the continuous bivariate distribution of the random fields $Z_1, Z_2$ it follows that $\mathrm{sgn}(Z_1(x) - Z_1(y)) \neq 0$ almost surely and similarly for $Z_2$. Therefore, $f(x,y)^2 = 1$ almost surely and from the boundedness assumptions on $\lambda_2,$ we obtain finite positive constants $m_3, M_3$ such that $m_3 |W|^2 \leq \mathbb{E}I_3 \leq M_3 |W|^2$.

Putting the results together and denoting $A=\mathbb{E}I_1 + 4 \mathbb{E}I_2, B = 2 \mathbb{E}I_3$ and $c_1 = (m_1 + 4 m_2)/\lambda^3, c_2 = (M_1 + 4 M_2) / \lambda^3, d_1 = 2 m_3 / \lambda^2, d_2 = 2 M_3 / \lambda^2$, we obtain the result for $\mathrm{var}(S)$.

Concerning the approximation to the variance of $S/U,$ we proceed as in the proof of Theorem~\ref{T1}, realizing that $\mathbb{E}S = 0$. It holds that $\mathbb{E}U = \int_{W^2} \lambda_2(u,v) \, \mathrm{d}u \, \mathrm{d}v,$ and the boundedness assumption on $\lambda_2$ now implies the claim for $\mathbb{E}U$. This concludes the proof.

\section{Variance correction factors -- simulation study}

\subsection{P-C test}\label{app:simulations_PC}
In this section, we provide empirical evidence supporting the choice of the correction factor suggested for the P-C test in Section~\ref{subsec:PC} for the test statistic $T$, i.e. the sample mean.

The variance of $T$ is of order $1/|W|$, see Theorem~\ref{T1}. Hence, $|W| \mathrm{var}(T)$ must be approximately constant as a function of $|W|$. To check this, we have performed the following simulation experiment.

Let $\Phi$ be a Poisson process with intensity 100. Let the covariate be given by a random field $Z(u)$, independent of $\Phi$. The random field $Z$ is centered, unit variance Gaussian random field with a spherical model for the correlation function. The scale parameter for the correlation function is chosen to be 0.05, 0.10, and 0.20, respectively.

\begin{figure}[tb]
    \centering
    \includegraphics[width=0.6\textwidth]{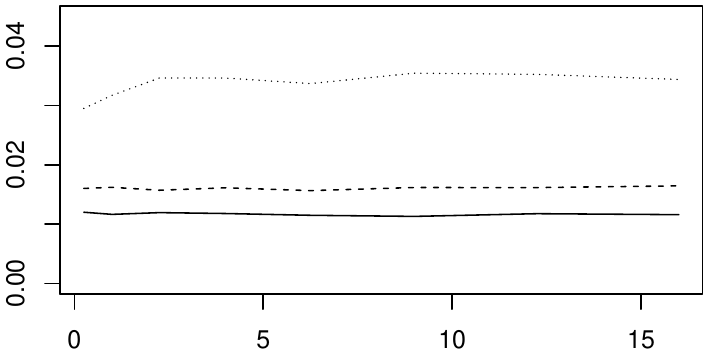}
    \caption{Sample variance of the sample mean multiplied by the area of the observation window (vertical axis) plotted against the area of the observation window (horizontal axis). The scale parameter for the correlation function of the random fields is chosen to be 0.05 (solid line), 0.10 (dashed line) and 0.20 (dotted line), respectively.}
    \label{fig:PC_variance}
\end{figure}

We consider a sequence of observation windows $[0,a]^2$ for $a=0.5,1,1.5, \ldots, 4$. For each observation window we generate $5\,000$ independent realizations of $(\Phi,Z)$ and compute the respective value of the test statistic $T$. We further compute the sample variance of these values and multiply it by the area of the observation window $|W| = a^2$. The resulting values are plotted in Figure~\ref{fig:PC_variance} as a function of $|W|$. For all three values of the scale parameter considered here the plotted functions are nearly constant. This indicates that the variance of the test statistic is, indeed, of order $1/|W|$.

\subsection{PM-C test}\label{app:simulations_PMC}

In this section, we provide empirical evidence supporting the choice of the correction factor suggested for the PM-C test in Section~\ref{subsec:PMC} for the test statistic $T_K$, i.e. Kendall's correlation coefficient. With this choice of the test statistic, the random shift test with variance correction performed well in the simulation studies in Section~\ref{sec:simulations}, even in situations with strong dependence between points and marks or covariate.

The variance of $T_K$ is of order $1/|W|$, see Theorem~\ref{T2}. Hence, $|W| \mathrm{var}(T_K)$ must be approximately constant as a function of $|W|$. To check this, we have performed the following simulation experiment.

Let $\Phi$ be a Poisson process with intensity 100. The marks follow the geostatistical marking model, i.e. they are obtained by sampling the values of a random field $M(u)$, independent of $\Phi$. Let the covariate be given by a random field $Z(u)$, independent of both $\Phi$ and $M$. The random fields $M,Z$ are identically distributed and are centered unit variance Gaussian random fields with a spherical model for the correlation function. The scale parameter for the correlation function is chosen to be 0.05, 0.10, and 0.20, respectively.

\begin{figure}[tb]
    \centering
    \includegraphics[width=0.49\textwidth]{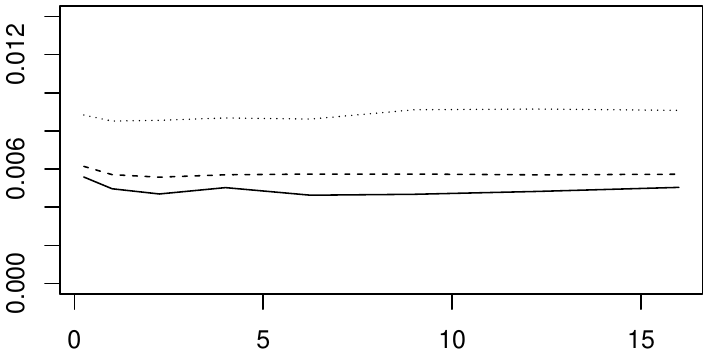}
    \includegraphics[width=0.49\textwidth]{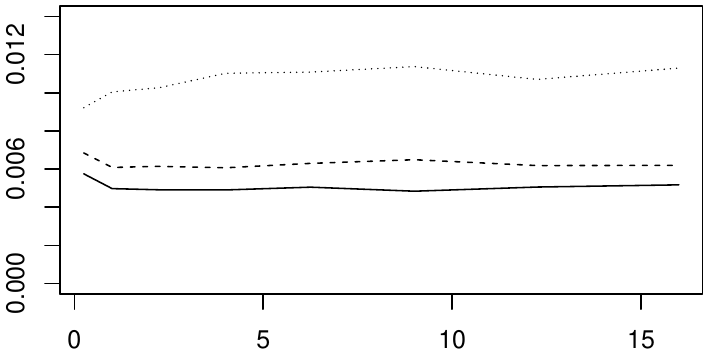}
    \caption{Sample variance of the Kendall's correlation coefficient multiplied by the area of the observation window (vertical axis) plotted against the area of the observation window (horizontal axis). The scale parameter for the correlation functions of the random fields is chosen to be 0.05 (solid line), 0.10 (dashed line) and 0.20 (dotted line), respectively. Left: both random fields have the same covariance function. Right: the two random fields have different covariance functions (spherical vs. exponential model), but with the same scale parameter.}
    \label{fig:PMC_variance}
\end{figure}

We consider a sequence of observation windows $[0,a]^2$ for $a=0.5,1,1.5, \ldots, 4$. For each observation window we generate $5\,000$ independent realizations of $(\Phi,M,Z)$ and compute the respective value of the test statistic $T_K$. We further compute the sample variance of these values and multiply it by the area of the observation window $|W| = a^2$. The resulting values are plotted in Figure~\ref{fig:PMC_variance} (left) as a function of $|W|$. For all three values of the scale parameter considered here the plotted functions are nearly constant. This indicates that the variance of the test statistic is, indeed, of order $1/|W|$.

To consider the more general situation of unequal covariance function, which is not covered by Theorem~\ref{T2}, we perform the same experiment with the mark field $M$ following the spherical model of correlation function again and the covariate $Z$ following the exponential model. The scale parameter for both correlation functions is again chosen to be 0.05, 0.10, and 0.20, respectively. The results are plotted in Figure~\ref{fig:PMC_variance} (right), with all three plotted curves nearly constant. This indicates that even in cases with different covariance functions (one of them having unbounded range), the variance of the test statistic is still of order $1/|W|$.

\end{document}